\documentclass[journal=jacsat,manuscript=article]{achemso}

\setkeys{acs}{doi=true}

\newcommand{\doi}[1]{\href{http://dx.doi.org/#1}{\nolinkurl{#1}}}

\usepackage[utf8]{inputenc}
\usepackage[english]{babel}
\usepackage{amsmath,amsfonts,amssymb}
\usepackage{graphicx}
\usepackage{hyperref}
\usepackage{indentfirst}
\usepackage{threeparttable}
\usepackage{multirow}
\usepackage{subcaption}
\usepackage{caption}
\usepackage{pdfpages}

\usepackage{orcidlink}

\usepackage{achemso}
\SectionNumbersOn

\newcommand\nnfootnote{%
  \begin{NoHyper}
  \renewcommand\thefootnote{}\footnote{Shared first authorship: 
BH and MCS contributed equally to this work}%
  \addtocounter{footnote}{-1}%
  \end{NoHyper}
}

\newcommand{\chfour}{CH$_4$\,}
\newcommand{\cotwo}{CO$_2$\,}
\newcommand{\fig}[1]{\textbf{Fig. (\ref{#1})}}
\newcommand{\eq}[1]{\textbf{Eq. (\ref{#1})}}
\newcommand{\avg}[1]{\langle{#1}\rangle}

\newcommand{\tab}[1]{\textbf{Table (\ref{#1})}}
\newcommand{\etal}{\emph{et al. }}
\newcommand{\hres}[2]{#1$^{(#2)}$}
\newcommand{\mtab}[2]{\textbf{Tables (\ref{#1}-\ref{#2})}}

\author{Basile Herzog\nnfootnote}
\affiliation[ULor]{Université de Lorraine and CNRS,
Laboratoire de Physique et Chimie Théorique, UMR 7019, 54506, 
Vand{\oe}uvre-lès-Nancy, France}

\author{Maurício Chagas da Silva\,\orcidlink{0000-0002-6890-0182}\nnfootnote}
\affiliation[ULor]{Université de Lorraine and CNRS,
Laboratoire de Physique et Chimie Théorique, UMR 7019, 54506, 
Vand{\oe}uvre-lès-Nancy, France}

\author{Bastien Casier\,\orcidlink{0000-0003-0261-4201}}
\affiliation[ULor]{Université de Lorraine and CNRS,
Laboratoire de Physique et Chimie Théorique, UMR 7019, 54506, 
Vand{\oe}uvre-lès-Nancy, France}

\author{Michael Badawi\,\orcidlink{0000-0002-3504-4180}}
\affiliation[ULor]{Université de Lorraine and CNRS,
Laboratoire de Physique et Chimie Théorique, UMR 7019, 54506, 
Vand{\oe}uvre-lès-Nancy, France}

\author{Fabien Pascale}
\affiliation[ULor]{Université de Lorraine and CNRS,
Laboratoire de Physique et Chimie Théorique, UMR 7019, 54506, 
Vand{\oe}uvre-lès-Nancy, France}

\author{Tomáš Bučko\,\orcidlink{0000-0002-5847-9478}}
\affiliation[TomA]{ Comenius University in Bratislava,
Department of Physical and Theoretical Chemistry, Faculty of
Natural Sciences, Mlynská Dolina, Ilkovičova 6, SK-84215
Bratislava, Slovakia}
\alsoaffiliation[TomB]{ Institute of Inorganic Chemistry, Slovak
Academy of Sciences, Dúbravská cesta 9, SK-84236 Bratislava,
Slovakia}
\email{tomas.bucko@uniba.sk}

\author{Sébastien Lebègue}
\affiliation[ULor]{Université de Lorraine and CNRS,
Laboratoire de Physique et Chimie Théorique, UMR 7019, 54506, 
Vand{\oe}uvre-lès-Nancy, France}

\author{Dario Rocca\,\orcidlink{0000-0003-2122-6933}}
\affiliation[ULor]{Université de Lorraine and CNRS,
Laboratoire de Physique et Chimie Théorique, UMR 7019, 54506, 
Vand{\oe}uvre-lès-Nancy, France}
\email{dario.rocca@univ-lorraine.fr}

\title{ Assessing the Accuracy of Machine Learning 
Thermodynamic Perturbation Theory: Density Functional 
Theory and Beyond }


\begin{document}

%
%

\begin{tocentry}

\centering
\includegraphics[width=8cm]{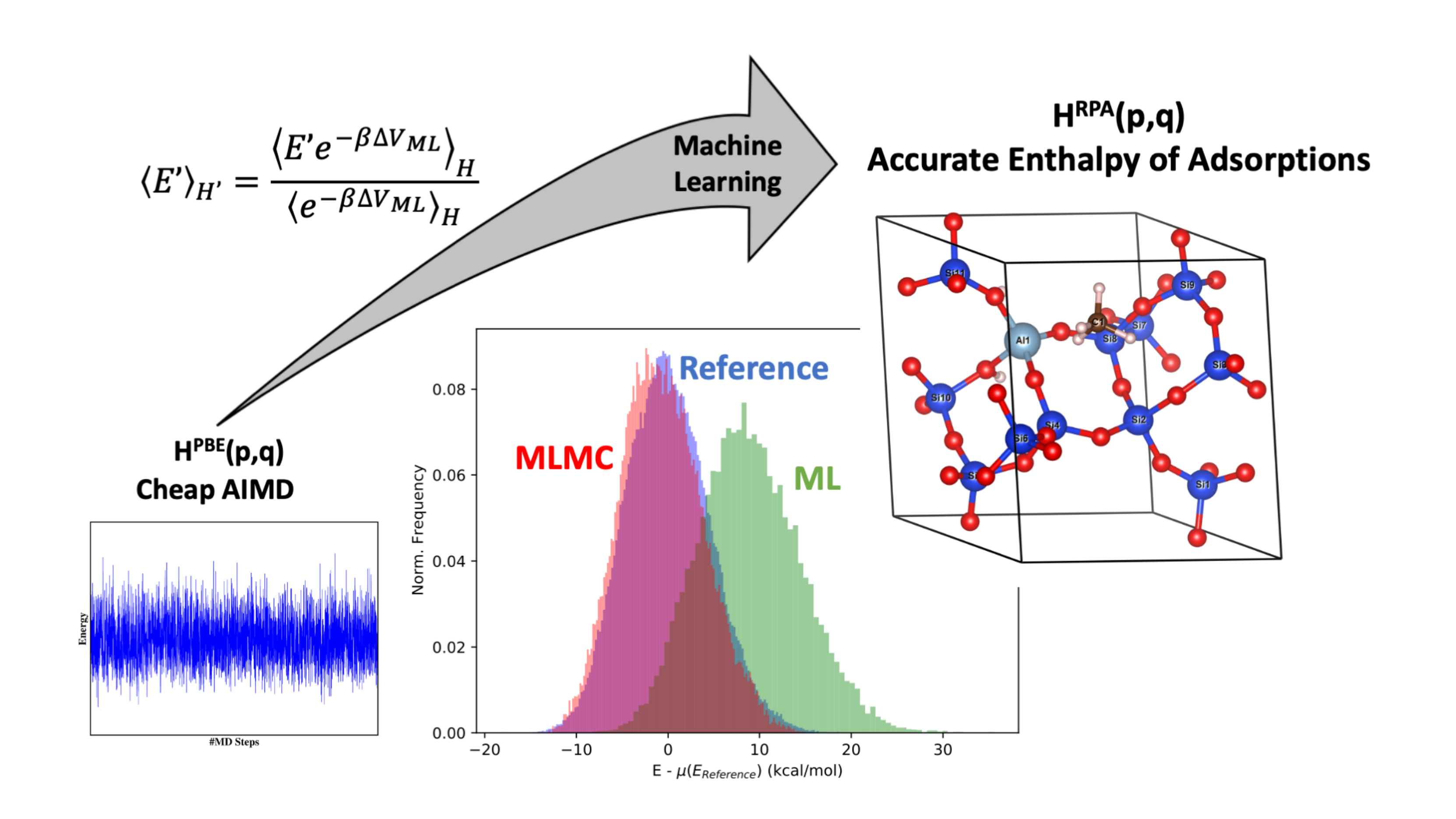}

\end{tocentry}

%
%

\clearpage

\begin{abstract}

  Machine learning thermodynamic perturbation theory (MLPT) is a promising
approach to compute finite temperature properties when the goal is to compare 
several different levels of \emph{ab initio} theory and/or to apply highly 
expensive computational methods. Indeed, starting from a production molecular 
dynamics trajectory, this method can estimate properties at one or more target 
levels of theory from only a small number of additional fixed-geometry 
calculations, which are used to train a machine learning model. However, as 
MLPT is based on thermodynamic perturbation theory (TPT), inaccuracies might
arise when the starting point trajectory samples a configurational space which 
has a small overlap with that of the target approximations of interest.
By considering case studies of molecules adsorbed in zeolites and several 
different density functional theory approximations, in this work we assess 
the accuracy of MLPT for ensemble total energies and enthalpies of adsorption. 
The problematic cases that were found are analyzed and it is shown that, even 
without knowing exact reference results, pathological cases for MLPT can be 
detected by considering a coefficient that measures the statistical imbalance 
induced by the TPT reweighting. For the most pathological examples we recover 
target level results within chemical accuracy by applying a machine 
learning-based Monte Carlo (MLMC) resampling.
Finally, based on the ideas developed in this work, we assess and confirm the 
accuracy of recently published MLPT-based enthalpies of adsorption at the random
phase approximation level, whose high computational cost would completely hinder 
a direct molecular dynamics simulation. 

\end{abstract}

%
%

\clearpage

\section{Introduction} \label{intro}

  Correlated quantum chemical methods could provide an alternative to
density functional theory (DFT) in (periodic) materials simulations
possibly reaching the threshold of chemical accuracy (1 kcal/mol)
with respect to experimental data. Traditional quantum chemical
methods, such as M\o ller-Plesset perturbation theory to second order
(MP2)~\cite{moller34} or coupled-cluster theory~\cite{bartlett07},
have been recently implemented for condensed phase materials
applications~\cite{pisani08,marsman09,delben2012,booth13,dixit2017communication}
. An alternative particularly suitable for condensed matter
applications is represented by the random phase approximation
(RPA)~\cite{bohm53,gell57,langreth75,dobson99,furche01,harl08,lu09}
and it variants which include higher order
corrections~\cite{gruneis09,olsen12,bates13,colonna14,dixit16,dixit17,hellgren18}. 
Due to the significantly high computational cost of these approaches,
their use is limited, especially in (finite-temperature) molecular dynamics (MD)
simulations. \\

The use of machine learning (ML) techniques could be highly beneficial for MD simulations
and, since the seminal work of Behler and Parrinello (BP) in 2007~\cite{behler07},
has seen an increasing popularity in this field~\cite{chmiela17,chmiela18,zhang18,behler16}.
While keeping a level of accuracy comparable to \emph{ab initio} calculations,
ML approaches can be used to replace most of the expensive
quantum mechanical calculations with numerically cheap predictions. This allows
for an increase in the system size and timescale that would be normally accessible
by traditional \emph{ab initio} MD simulations. However, ML models typically
employed in MD simulations require a significant amount of data to be trained
and this represent an issue for the most expensive quantum mechanical approximations. \\

  In a recent work, some of us applied a scheme that couples machine learning
techniques with thermodynamic perturbation theory (MLPT) to compute enthalpies
of adsorption of molecules in the zeolite chabazite at the RPA level of
theory~\cite{Chehaibou2019} (the same methodology can be used in the context
of free energies of activation~\cite{Bucko2020,Gesvandtnerova2021}). Achieving
these results with a brute force molecular dynamics simulation would be
completely out of reach. Indeed, by considering only the cost of the RPA energy
calculations for the zeolite, completing a 200000 steps MD simulation would
require 65M CPU hours and more then 110 years on 64 cores at 2.6 GHz 
(this estimate does not include
the significant additional cost of computing RPA forces~\cite{ramberger17}).
With our MLPT technique as few as 10 single point RPA energy calculations were
sufficient to train a machine learning model that was then used to predict the
RPA energy for several other configurations~\cite{Chehaibou2019}.\\

  In this previous work the van der Waals (vdW) corrected PBE+D2 functional
~\cite{Grimme2006,Bucko2010} was used to generate a production MD trajectory
(in general, a numerically inexpensive approximation should be chosen as
production method). Based on the $\Delta$-ML method~\cite{Ramakrishnan2015},
a model was thereafter trained to predict the difference between RPA and PBE+D2 energies. 
Following the thermodynamic perturbation approach 
~\cite{chipot07,Pohorille2010,Rocca2019} the energy differences were subsequently
used to reweight the statistical weight of the PBE+D2 configurations in order
to obtain the RPA canonical distribution and compute the RPA ensemble energies
and enthalpies (in this context the RPA is considered the ``target'' level of
theory). Thanks to the inexpensive predictions of the ML model, this technique
involves a computational cost that is several orders of magnitude smaller than a
full RPA MD simulation.\\

  While the error involved in the ML procedure is rather well controlled
~\cite{Chehaibou2019,Bucko2020}, the application of thermodynamic perturbation
theory (TPT) might be at the origin of a bias in the results estimated
at the target level of theory. Indeed, the configurational space sampled
by the production MD might have suboptimal superposition with the target
configurational space~\cite{chipot07,Pohorille2010}. In certain cases this might
lead to a strong loss of statistical significance when the contributions of
the production configurations are reweighted to obtain the target distribution
(namely only few configurations could contribute to the whole target level
statistics).\\

  In this work, we access the accuracy of the MLPT approach using the adsorption
of molecules in zeolites as a test case. Indeed, many separation and adsorption 
processes require the improvement of the dedicated materials, and DFT calculations 
often provide suitable suggestions of optimized formulations in this regard
~\cite{Adil2017,Khalil2020,Chibani2016,Hessou2021}. 
However, additional efforts have to be done to find a better compromise between 
accuracy of prediction and calculation cost. By considering five different DFT
functionals, including generalized gradient approximation (GGA), meta-GGA,
and non-local vdW corrected functionals, we perform full MD simulations to
provide ``exact'' references values for the ensemble energies and enthalpies
of adsorption. We then use each one of these five functionals as a production
method and apply MLPT to obtain the energy/enthalpy estimates for the remaining
four functionals, which are considered as target approximations.\\

  By comparing MLPT estimates with exact reference results the level of accuracy
of this approach is established. Some pathological cases are found, for example
when GGA functionals are used as starting point for meta-GGA functionals and
vice versa (in certain cases the MLPT estimate of the ensemble energy can
deviate by more than 8 kcal/mol). By using machine learning dimensionality
reduction algorithms, we qualitatively visualize the relative distributions
of the configurations from the different MD simulations. This analysis confirms
that the most problematic cases can be ascribed to a poor superposition of the
configurational spaces. In order to detect possible failure of MLPT we propose a
diagnostic test, the $I_w$ index, which is evaluated from ML energy predictions
and does not rely on the knowledge of any exact reference results.\\

  We then propose a scheme to significantly improve results even when the $I_w$
coefficient is close to 0 (lowest production-target superposition). This
approach, denoted as MLMC, is based on a Monte Carlo resampling of the target
configurational space that reuse the machine learning model already trained for
MLPT. Without any additional target level calculations this scheme decreases
even the largest deviations within the threshold of chemical accuracy.\\

  Finally, we apply the tools developed in this work to analyze our previous
results based on the RPA~\cite{Chehaibou2019}, where reference values cannot
be produced. The relatively large values of the $I_w$ index hints that the
PBE+D2 functional provides a reliable starting point for RPA target properties.
The results are stable even if a full MLMC resampling of the configurational
space is performed. This shows that, with a proper choice or the production
approximation, MLPT allows for a quick and accurate estimate of target level
properties. The MLPT approach opens the way to a more systematic application
of accurate but expensive DFT and quantum chemical approximations in finite
temperature simulations of materials.\\

  The layout of the paper is organized as follows. In the following section,
Sec. \ref{sec:methodology}, we discuss the methodological part of this work
and provide computational details. In Sec. \ref{results}, we present numerical
results and analyze them. Sec. \ref{conclusions} contains our conclusions.

%
%

\clearpage

\section{Methodology}\label{sec:methodology}

\begin{figure}[hbtp!] 
  \includegraphics[width=12cm]{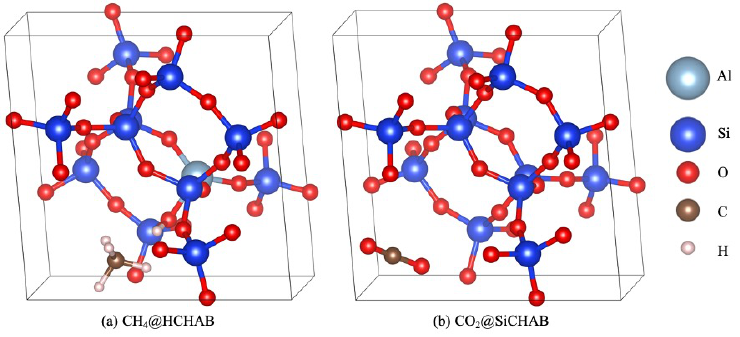}
  \caption{(a) \chfour in protonated Chabazite (HCHAB) and (b) \cotwo in siliceous
  chabazite (SiCHAB).} 
  \label{fig:mysys} 
\end{figure}

   The enthalpies of adsorption($\Delta_{ads}H$) of the \chfour
molecule in protonated chabazite (HCHAB) and \cotwo molecule in siliceous chabazite 
(SiCHAB) were investigated by employing \emph{ab initio}
molecular dynamics (AIMD) and machine learning thermodynamic perturbation 
theory~\cite{Chehaibou2019,Bucko2020,Gesvandtnerova2021} (MLPT). The models 
of the adsorbed molecules are shown in \fig{fig:mysys}. The enthalpy of adsorption 
is defined by 

\begin{equation}
  \Delta_{ads}H(M@Z) = \avg{E(M@Z)} - ( \avg{E(M)} + \avg{E(Z)} ) - k_{B}T
  \label{eq:dH}
\end{equation}   

where $\avg{E}$ is the internal energy computed as ensemble average of potential
energy via the AIMD simulations or MLPT (M denotes the molecules, Z the
zeolites, and M@Z the adsorbed system), k$_{B}$ is the Boltzmann constant, and
$T$ is the system temperature (equal to 300K in all our simulations).\\

  The goal of this work is to benchmark the accuracy of MLPT by comparing
its estimates with ``exact'' reference values obtained from full MD
simulations. We chose five DFT functionals in order to span a range of
different characteristics but keeping also into account their numerical
cost, as long and stable MD trajectories have to be produced for reference.
Specifically, we selected: a generalized gradient approximation (GGA)
functional, PBE~\cite{Perdew1996} and its version with corrections for van
der Waals interactions, PBE+D2~\cite{Grimme2006,Bucko2010}; a meta-GGA
functional, SCAN~\cite{Sun2015,Sun2016}; a non-local vdW functional,
vdW-DF2~\cite{Roman-Perez2009,Dion2004}; a version of SCAN with non-local vdW
corrections, SCAN+rVV10~\cite{Peng2016}. We will also discuss results based
on the random phase approximation (RPA), as implemented in the VASP code
~\cite{Kresse1993,Kresse1996,Kresse1996a}; due to the high numerical cost a
direct MD simulation at this level of theory is completely out of reach and the
MLPT method becomes instrumental to estimate finite temperature properties.\\

  All AIMD simulations were performed within the NVT ensemble, whereby
temperature was maintained by Andersen~\cite{Andersen1980} thermostat with
collision probability of 0.05. An integration time step of 0.5 fs was
considered and the total simulation time of each MD run was 100 ps, i.e.
$\sim$200$\cdot 10^3$ configurations were generated with each method for each
system investigated in this work. The initial part of 10 ps of each trajectory
was considered as an equilibration period and the corresponding data were
discarded. The remaining part of the trajectory was tested for the absence of any
drift via Mann-Kendall tests~\cite{Schiferl1985}. All the standard errors on 
the ensemble energies of these trajectories are below 0.2 kcal/mol.\\

  In the AIMD simulations we used a cell fixed at the lattice parameters
of the chabazite optimized at the PBE level ($a=b=c=9.34$ \AA\, and
$\alpha=\beta=\gamma=95^{o}$). All AIMD
simulations as well as single point calculations were carried out with the
VASP electronic structure package~\cite{Kresse1993,Kresse1996,Kresse1996a,harl08}.
PAW pseudopotentials~\cite{Kresse1999} for all atoms with the default kinetic
energy cutoffs were used to expand the wavefunctions. The ``accurate'' precision
setting was used and the hydrogen atomic mass was increased to 3.0 a.u. The
$\Gamma$ point approximation was applied in all calculations. \\

  The results produced by AIMD are used as a reference for our MLPT method. Within
this approach an MD simulation is first performed at a certain production
level of theory and then ``corrected'' to obtain results at one or multiple
target levels of theory. One of the advantages of MLPT is that an estimate of a
finite-temperature property for the target method can be obtained with only a
few tens or few hundreds of single point calculations~\cite{Chehaibou2019}. Out
of about 1900$\cdot 10^3$ configurations generated in production runs, a reduced
sample of about 19$\cdot 10^3$ evenly separated configurations (${X_i}$) and
corresponding potential energies obtained with the production method (${E_i}$)
are selected to be used for the predictions of the energies at the new target
functional level. We note that this configuration reduction is not strictly
necessary but, since MD configurations close in time are correlated, this
procedure speeds up the predictions of the ML model without compromising the
accuracy of prediction. \\

  In order to compute target level ensemble energies $\avg{E'}$, MLPT applies
TPT to reweight the production statistical distribution~\cite{Pohorille2010}

\begin{equation} 
  \avg{E'} = \frac{\sum_{i}^{M} E_i' \exp \left( -\beta \Delta E_i \right)}{\sum_{i}^{M} \exp \left( -\beta \Delta E_i \right)} 
  \label{eq:rew}
\end{equation} 

where $E_i'$ denotes the target level energies of the $i$th
configuration, $\Delta E_i=E_i'-E_i$ is the difference between target and
production energies, $\beta = 1/k_BT$, and $M$ is the number of configurations,
in this case 19$\cdot 10^3$.\\

  The application of \eq{eq:rew} still requires a large number ($M$) of
calculations at the target level of theory. In MLPT only a very small number
$N_{train}$ of these calculations is actually performed and then used to train
an ML model that predicts all the remaining $M-N_{train}$ energies. This
procedure has been described in detail in Ref.~\cite{Chehaibou2019} and here we
will only summarize the main technical points. In order to decrease the number
of target calculations required to train the ML algorithm instead of considering
directly the target energy $E_i'$, we build a ML regression model using $\Delta
E_i$. This approach, which is just one of the many possible applications of
the $\Delta-$ML idea~\cite{Ramakrishnan2015,Deringer2021}, takes advantage
of the much smoother dependence of $\Delta E_i$ on $E_i$ in comparison with
$E_i'$ and, accordingly, is easier to predict. To represent the configurations
within our ML model we use the \emph{Smooth Overlap of Atomic Positions}
(SOAP)~\cite{Bartok2013} descriptor, as implemented in the DScribe library
~\cite{Himanen2020}. While several other descriptors for periodic materials have
been proposed in the literature~\cite{behler11,huo17,Casier2021}, the choice of
the SOAP descriptor provides already a satisfactory level of accuracy for MLPT
applications~\cite{Chehaibou2019,Bucko2020,Gesvandtnerova2021}.\\

  The SOAP approach leads to a natural definition of a kernel for local
atomic environments that can be subsequently used to define the global
rematch kernel~\cite{Stuke2019}. This kernel, which can be intuitively
seen as a measure of similarity between configurations, is then used in
the framework of the kernel ridged regression (KRR)~\cite{Stuke2019} ML
algorithm; the KRR implementation in the Scikit-learning package is used in
this work~\cite{Pedregosa}. To train the ML model, 200 evenly distributed
configurations are chosen from the 19$\cdot 10^3$ structures of the production
trajectory. An additional independent set of 25 configurations is also
selected to test the accuracy of the predictions of the ML model; 
for all the applications considered in this work the root mean square 
error (RMSE) in the prediction of the energy of single configurations 
is at most 0.3 kcal/mol. Single point
calculations at the target level are performed exclusively for these 225
configurations. The 18775 (i.e., 19000-225) remaining values of $\Delta E_i$
(and $E_i'=E_i+\Delta E_i$) necessary to evaluate \eq{eq:rew} are inexpensively
predicted by the ML model. This shows the clear advantage of using MLPT, in
particular if it is necessary to consider multiple target level theories or
highly expensive approximations (e.g., the RPA). \\

  As discussed in the next sections MLPT does not provide satisfactory results
in certain cases. This happens for the estimate of SCAN and SCAN+rVV10
adsorption enthalpies from PBE production calculations.
This issue is related to the unsatisfactory superposition of the target
and production configurational spaces. To improve the results in these
cases we have reused the ML model from the MLPT application to resample
the target level configurational space using the Monte Carlo (MC)
algorithm~\cite{Paquet2015,Kremer1991,Meimaroglou2014,Betancourt2019}. We will
denote this procedure as MLMC.\\

  The MC approach used to sample the canonical ensemble is equivalent to MD but
is simpler to implement and does not require the calculation of forces (for
certain methods forces are not available in most solid state
physics software implementations). By using our in-house software, the MLMC
approach recursively sample new configurations following these steps: (1) A new
geometric configuration is randomly generated; (2) A production level \emph{ab initio} 
calculation is performed to obtain $E_i$ for this new configuration;
(3) the ML learning model is used to predict $\Delta E_i$ and $E_i'$; (4)
the new configuration is accepted or rejected according to the Metropolis
criterion~\cite{Paquet2015,Kremer1991,Meimaroglou2014,Betancourt2019}. The
advantage of MLMC is that with respect to MLPT no additional calculations are
required at the target level of theory and this is particularly important when
computationally expensive quantum chemical methods are used. The trade-off is
that each step of the MC procedure requires a new production level calculation.
In the applications of next section we have sampled 400$\cdot 10^3$ steps for
each MLMC trajectory to get converged ensemble statistics (for all the systems 
considered in this work the standard error on the MC ensemble energies is at
most 0.3 kcal/mol).

%
%

\clearpage

\section{Results and Discussions}\label{results}

\subsection{Assessing the accuracy of MLPT for different density functional
approximations}

  The experimentally measured enthalpies of adsorption of the investigated
systems are rather small, -4.06 kcal/mol and -5.02 kcal/mol for \chfour @ HCHAB
and \cotwo @ SiCHAB, respectively. Indeed, for both systems, the adsorption
process is dominated by weak long-range van der Waals (vdW) interactions,
which are, due to the small size of both adsorbate molecules, only modest.
Ideally, very accurate methods such as RPA, MP2 or CCSD would be necessary
to model with systematic improvements in accuracy such processes in which
long-range interactions play an important role. While the use of such methods
is prohibitive due to their enormous computational cost, machine learning
techniques could play an important role in extending their applicability to
realistic systems~\cite{Chehaibou2019}. This requires ML techniques that
reasonably preserve the accuracy of these high-level quantum chemical methods
and have sufficient predictive power for configurations beyond the training
set. In this section we will discuss the accuracy of MLPT for the case of
computationally affordable DFT functionals, for which an ``exact'' reference
can be easily obtained. At the end of this section we will consider the RPA
enthalpies of adsorption.\\

  In \tab{tab:dHexp}, we report the values of the enthalpies of adsorption (\eq{eq:dH})
for \chfour and \cotwo in zeolite as obtained from the full MD
simulations. These results will be used as reference values for assessing the
quality of our MLPT methodology. Comparing the calculated and the experimental
values for the enthalpies of adsorption in \tab{tab:dHexp}, it can be noticed
that only functionals with some vdW corrections can achieve a reasonable agreement
with the experimental values, although the accuracy improvement is not
systematic. For the \chfour@HCHAB system, SCAN+rVV10 was identified as the best
performing functional, with a deviation of -0.71 kcal/mol (17\% of deviation),
whereas for the \cotwo@SiCHAB the best functional was PBE+D2 with a deviation of
-0.46 kcal/mol (9\% of deviation). On the other hand, the uncorrected PBE and
SCAN functionals show a sizable underestimation in the calculated enthalpies of
adsorption for the \chfour and \cotwo molecules in the zeolites, with relative
errors with respect to experiment higher than 35\%. However, the main purpose of
this work is not to compare the accuracy of different functionals compared to
experiment but to assess the performance and limitations of MLPT to reproduce
exact AIMD results obtained using different quantum mechanical approximations.\\

\begin{table}[htpb!]
\centering
\caption{\label{tab:dHexp} Experimental and reference theoretical enthalpies 
of adsorption for the \chfour and \cotwo molecules in zeolites obtained using 
different functionals. All values are expressed
in kcal/mol.}
\begin{tabular}{ccccccc}
\hline
\hline
\\
System         & PBE   & PBE+D2 &  SCAN & vdW-DF2 & SCAN+rVV10 & Exp. \\
\hline
\\
\chfour@HCHAB  & -1.13 & -6.08  & -2.60 & -5.64 & -4.77 & -4.06~\cite{Piccini2015}\\
\cotwo@SiCHAB  & -1.18 & -5.48  & -2.95 & -6.93 & -7.11 & -5.02~\cite{Maghsoudi2013}\\
\\
\hline
\hline
\end{tabular}
\end{table}

  Starting from each of the 5 trajectories obtained using PBE, PBE+D2, SCAN,
vdW-DF2, and SCAN+rVV10 production methods, the MLPT approach is used to
predict the results for all remaining approximations. The results are shown
in \tab{tab:dHRew}. In this table, columns and lines represent the production
and target methods, respectively. If, for example, we consider the PBE column
(below Production - MD simulations), and the SCAN line, we find an estimate of
the SCAN enthalpy of adsorption that is obtained by applying MLPT to the PBE
production trajectory; this estimate avoids completely the generation of a
new SCAN molecular dynamics trajectory and requires only 200 additional SCAN
calculations that are used to train the ML model. For the sake of completeness,
we also report values on the diagonal of this table, namely MLPT values based
on the same production and target levels of theory. This shows that some small
fluctuations can be introduced by the ML model even in this particularly simple
case, additionally to those regular/general fluctuations due to the AIMD
convergence and statistical uncertainty.\\

  The accuracy of the MLPT estimates can be established by comparing the MLPT
results with reference values of \tab{tab:dHexp}. The corresponding deviations
are reported in parenthesis in \tab{tab:dHRew}. Altogether, it can be noticed
that the predictions within the PBE/PBE+D2/vdW-DF2 group and the SCAN/SCAN+rVV10
are reasonable with deviations typically within few tens of kcal/mol and only
in one case slightly beyond chemical accuracy (a deviation of 1.01 kcal/mol for
the prediction of SCAN+rVV10 from SCAN for \cotwo@SiCHAB). When the production
and target methods belong to two different groups of functional, the MLPT is
significantly less reliable and significantly large deviations appear. The most
problematic cases involve the prediction of \chfour@HCHAB enthalpy of adsorption
at the SCAN and SCAN+rVV10 levels of theory from PBE, where the deviation is
larger then 8 kcal/mol.\\

\begin{table}[htpb!]
\footnotesize
\centering
\caption{\label{tab:dHRew} Enthalpies of adsorption of 
\chfour@HCHAB and \cotwo@SiCHAB computed directly from straightforward 
MD (Ref. column) as well as using the MLPT method, whereby all functionals have 
been used as production as well as target methods. Deviations of MLPT results 
from the reference values (Ref. column) 
are given in parenthesis. All values are expressed in kcal/mol.}
\begin{tabular}{cccccccc}
\hline
\hline
       &        & \multicolumn{5}{c}{Production - MD simulations} \\
\cline{4-8}
System & Target - MLPT & Ref. & PBE & PBE+D2 & SCAN & vdW-DF2 & SCAN+rVV10 \\
\hline
\\
              &         PBE & -1.13 & \hres{ -1.18}{-0.05} & \hres{ -1.05}{ 0.07} & \hres{  1.25}{ 2.38} & \hres{ -0.66}{ 0.47} & \hres{  0.23}{ 1.36} \\
              &      PBE+D2 & -6.08 & \hres{ -5.99}{ 0.09} & \hres{ -6.08}{ 0.00} & \hres{ -2.64}{ 3.44} & \hres{ -5.10}{ 0.98} & \hres{ -5.65}{ 0.43} \\
\chfour@HCHAB &        SCAN & -2.60 & \hres{-11.13}{-8.54} & \hres{ -4.78}{-2.18} & \hres{ -2.65}{-0.05} & \hres{ -1.84}{ 0.75} & \hres{ -2.37}{ 0.23} \\
 Exp. -4.06~\cite{Piccini2015} & vdW-DF2 & -5.64 & \hres{ -6.24}{-0.59} & \hres{ -5.82}{-0.18} & \hres{ -0.92}{ 4.73} & \hres{ -5.53}{ 0.11} & \hres{ -5.75}{-0.11} \\
             &  SCAN+rVV10 & -4.77 & \hres{-13.63}{-8.86} & \hres{ -6.71}{-1.94} & \hres{ -5.03}{-0.26} & \hres{ -4.13}{ 0.64} & \hres{ -4.74}{ 0.03} \\
\\
              &         PBE & -1.18 & \hres{ -1.15}{ 0.02} & \hres{ -1.47}{-0.29} & \hres{ -1.46}{-0.29} & \hres{ -1.24}{-0.06} & \hres{ -4.68}{-3.50} \\
              &      PBE+D2 & -5.48 & \hres{ -5.61}{-0.14} & \hres{ -5.44}{ 0.04} & \hres{ -5.40}{ 0.08} & \hres{ -5.60}{-0.12} & \hres{ -7.56}{-2.08} \\
\cotwo@SiCHAB &        SCAN & -2.95 & \hres{ -1.42}{ 1.54} & \hres{ -2.74}{ 0.21} & \hres{ -2.96}{-0.01} & \hres{ -4.30}{-1.34} & \hres{ -3.72}{-0.77} \\
 Exp. -5.02~\cite{Maghsoudi2013} & vdW-DF2 & -6.93 & \hres{ -7.47}{-0.54} & \hres{ -7.57}{-0.64} & \hres{ -8.11}{-1.17} & \hres{ -6.94}{-0.00} & \hres{ -8.90}{-1.96} \\
              &  SCAN+rVV10 & -7.11 & \hres{ -3.97}{ 3.14} & \hres{ -5.52}{ 1.59} & \hres{ -6.01}{ 1.10} & \hres{ -7.17}{-0.06} & \hres{ -7.09}{ 0.02} \\
\\
\hline
\hline
\end{tabular}
\end{table}
\normalsize

\clearpage

  As the enthalpies of adsorption are computed from differences of internal
energies of interacting and non-interacting systems, fortuitous error
cancellations might arise that could hide the real accuracy of MLPT predictions.
To shed some light onto this problem, we report in \tab{tab:ergrew} MLPT total
ensemble energies for the individual components involved in the adsorption
process (adsorbed system, zeolite alone, molecule alone). These results confirm
the conclusions previously drawn from \tab{tab:dHRew}, with the predictions made
for the cases where the production and target methods are both from the same
group of functionals (i.e., either PBE/PBE+D2/vdW-DF2 or SCAN/SCAN+rVV10) being
well within the chemical accuracy. \\

  MLPT becomes unreliable when predictions mix these two groups, with
deviations that often reach several kcal/mol (these observations do not hold
for the standalone \chfour and \cotwo molecules, whose deviations are always
below 0.1 kcal/mol). The worst performance is confirmed to correspond to
the prediction of SCAN and SCAN+rVV10 from the PBE production trajectory of
\chfour@HCHAB. For these cases, we notice that the largest deviations
in the enthalpies of adsorption ($>$8 kcal/mol) are not only due to the low
accuracy of the MLPT ensemble energy estimates for \chfour@HCHAB and HCHAB,
but also to the opposite sign in deviations of predictions made for these two
systems, which leads to an error accumulation when evaluating \eq{eq:dH}.\\

 While MLPT allows, in principle, for a quick evaluation of finite-temperature
properties at one or more target levels of theory from a single MD production
run (typically based on the most computationally inexpensive approximation),
the results in \mtab{tab:dHRew}{tab:ergrew} show that such a strategy does not
always allow one to achieve the required level of accuracy. Accordingly, a
special care should be taken in choosing a production method suitable for the
target approximation(s) of interest. The failure of MLPT in certain cases can
be explained by the limitations of the thermodynamic perturbation theory (TPT)
itself, rather than by an inaccurate ML model. Indeed, if the high-probability
configurational spaces of the production and target approximations do not
overlap sufficiently, the TPT has low predictive power~\cite{Pohorille2010}.
In the most pathological cases it can happen that this overlap is so poor
that in the reweighting procedure (\eq{eq:rew}) only one or few individual
configurations effectively contribute to the ensemble average. \\

\begin{table}[htpb!]
\footnotesize
\centering
\caption{\label{tab:ergrew} Deviations of the MLPT estimates of the target internal energies
of individual systems 
from reference values.
All values are  in kcal/mol.}
\begin{tabular}{ccccccc}
\hline
\hline
       &        & \multicolumn{5}{c}{Production - MD simulations} \\
\cline{3-7}
System & Target - MLPT & PBE & PBE+D2 & SCAN & vdW-DF2 & SCAN+rVV10 \\
\hline
\\
              &        PBE & -0.007 &  0.053 & -1.348 & -0.237 & -1.582 \\
              &     PBE+D2 &  0.197 &  0.001 & -0.547 &  0.405 & -2.987 \\
\chfour@HCHAB &       SCAN & -4.380 & -2.137 & -0.029 &  1.497 &  0.204 \\
              &    vdW-DF2 & -0.378 &  0.206 &  0.285 &  0.060 & -3.799 \\
              & SCAN+rVV10 & -5.026 & -2.036 & -0.221 &  1.012 & -0.001 \\
\\
              &        PBE &  0.014 & -0.693 & -2.946 & -0.434 & -3.689 \\
              &     PBE+D2 &  0.245 &  0.041 & -2.267 &  0.015 & -1.665 \\
\cotwo@SiCHAB &       SCAN &  2.249 &  1.244 & -0.013 &  2.448 & -0.594 \\
              &    vdW-DF2 & -0.053 & -0.622 & -4.138 & -0.002 & -2.122 \\
              & SCAN+rVV10 &  3.395 &  2.289 &  0.899 &  3.251 &  0.025 \\
\\
              &        PBE &  0.035 &  0.025 & -3.706 & -0.712 & -2.953 \\
              &     PBE+D2 &  0.054 &  0.000 & -4.007 & -0.626 & -3.479 \\
HCHAB         &       SCAN &  4.142 &  0.084 &  0.025 &  0.718 & -0.034 \\
              &    vdW-DF2 &  0.230 &  0.453 & -4.407 & -0.039 & -3.678 \\
              & SCAN+rVV10 &  3.830 & -0.045 &  0.051 &  0.359 & -0.027 \\
\\
              &        PBE & -0.012 & -0.380 & -2.634 & -0.348 & -0.230 \\
              &     PBE+D2 &  0.360 &  0.001 & -2.337 &  0.145 &  0.353 \\
SiCHAB        &       SCAN &  0.659 &  1.050 &  0.002 &  3.809 &  0.139 \\
              &    vdW-DF2 &  0.453 &  0.008 & -2.972 &  0.001 & -0.228 \\
              & SCAN+rVV10 &  0.246 &  0.745 & -0.158 &  3.358 &  0.000 \\
\\
              &        PBE &  0.006 & -0.047 & -0.024 &  0.009 &  0.012 \\
              &     PBE+D2 &  0.053 &  0.000 &  0.021 &  0.053 &  0.062 \\
\chfour       &       SCAN &  0.013 & -0.039 & -0.001 &  0.027 &  0.012 \\
              &    vdW-DF2 & -0.014 & -0.067 & -0.033 & -0.007 & -0.014 \\
              & SCAN+rVV10 &  0.000 & -0.052 & -0.013 &  0.014 & -0.000 \\
\\
              &        PBE &  0.003 & -0.018 & -0.026 & -0.025 &  0.041 \\
              &     PBE+D2 &  0.021 & -0.000 & -0.007 & -0.007 &  0.060 \\
\cotwo        &       SCAN &  0.054 & -0.018 & -0.006 & -0.018 &  0.037 \\
              &    vdW-DF2 &  0.030 &  0.007 &  0.004 & -0.001 &  0.070 \\
              & SCAN+rVV10 &  0.013 & -0.049 & -0.043 & -0.051 &  0.001 \\

\\
\hline
\hline
\end{tabular}
\end{table}

\clearpage

\begin{figure}[htbp!]
\includegraphics[width=16cm]{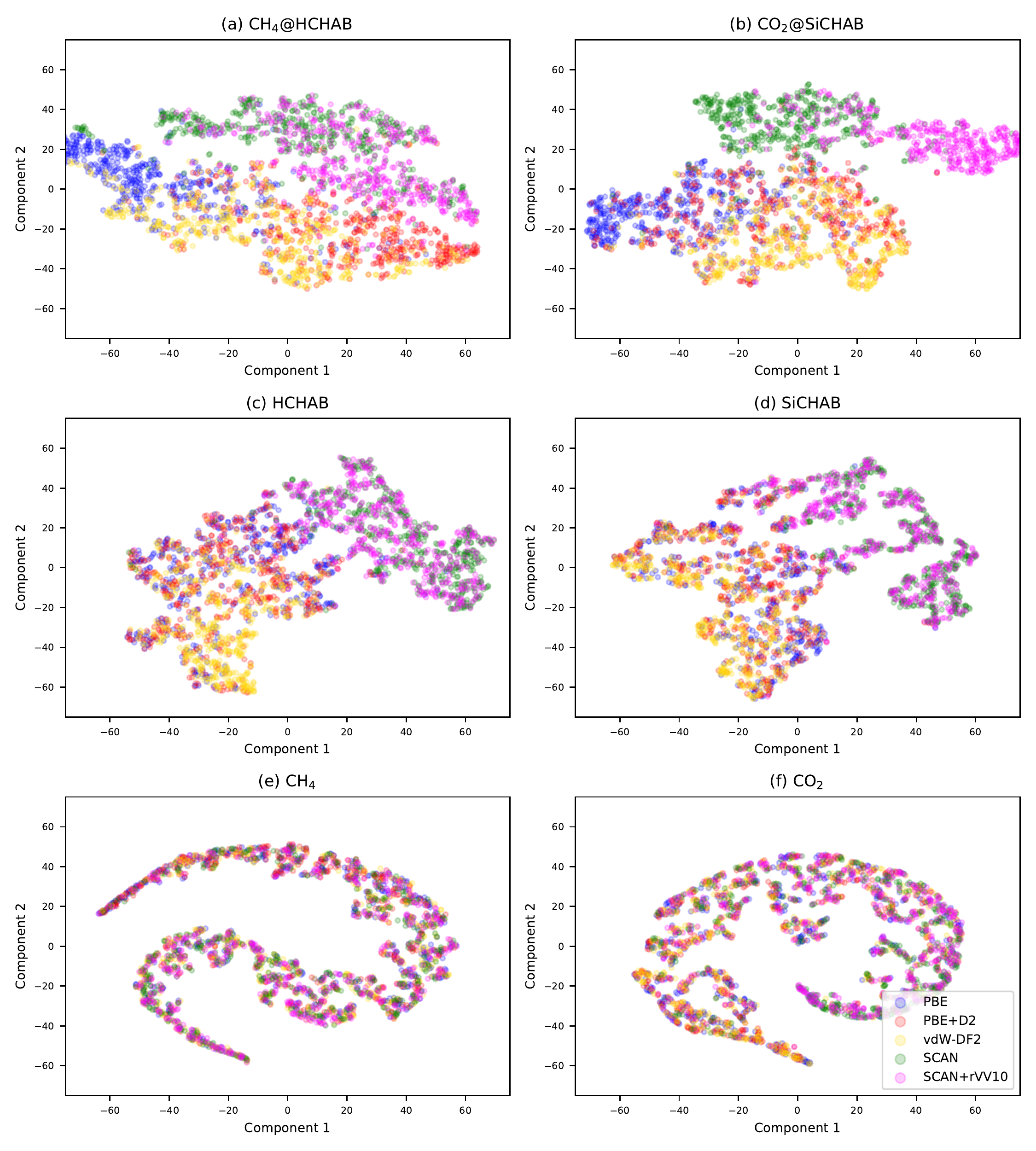}
\caption{t-SNE representation of the configurational spaces spanned
by the different functionals. This visualization is based on 500 selected 
configurations from each molecular dynamics trajectory.
principal component of the
SOAP descriptor for each 500 configurations of each functionals}
\label{fig:tsne}
\end{figure}

  In order to analyze qualitatively this behavior we consider all the production
MD simulations based on the 5 different functionals. By using the t-distributed
stochastic neighbor embedding (t-SNE) algorithm~\cite{VanDerMaaten2008,De2016}
we present in \fig{fig:tsne} two-dimensional visualizations of the
high-dimensional configurational space spanned by these trajectories (to improve
the readability, figures were created using only 500 uncorrelated structures
selected from each trajectory). To be consistent with the ML learning algorithm
used in this study for the regression, the t-SNE approach was applied using a
definition of distance $D$ between two configurations $\chi^A$ and $\chi^B$
based on the normalized SOAP kernel $K$:

\begin{equation} 
  D(\chi^A,\chi^B) = (1-K(\chi^A,\chi^B))^{\frac{1}{2}} ;
  \label{eq:tsne} 
\end{equation} 

this definition also inherits some of the properties of the SOAP kernel, such as the
rotational and translational invariance required in materials and molecular
modelling.\\

  It can be noticed in \fig{fig:tsne} that for \chfour@HCHAB, \cotwo@SiCHAB,
HCHAB, and SiCHAB the trajectories generated by the two groups of functionals
(PBE/PBE+D2/vdW-DF2 and SCAN/SCAN+rVV10) form two clusters with very limited
overlap. This shows that the configurational spaces spanned by the two groups
of approximations are to a large extent different, which is at the origin
of the poor performance of MLPT in the cases discussed above. For the gas
phase molecules \chfour and \cotwo, the configurational space has a much
simpler structure and the five different approximations produce trajectories
which always largely overlap; this is also consistent with the results in
\tab{tab:ergrew}, where MLPT is highly accurate for all molecular energies.\\

\clearpage

\clearpage

  The t-SNE visualization provides useful insights on the potential failures of
TPT and, consequently, MLPT. However, a diagnostic approach that requires the
full MD trajectories is highly impractical in reality, since the goal of MLPT
is indeed to completely avoid the MD simulation at the target level of theory. In
order to overcome this difficulty and introduce a tool to analyze and predict
the failures of MLPT, we follow the approach introduced in Ref.~\cite{Bucko2020}
and introduce the $I_w$ index. By considering the weight factor ($w_i$) in
\eq{eq:rew}

\begin{equation} 
  w_i = exp(-\beta \Delta E_i) 
  \label{eq:wgh} 
\end{equation}

we define $ I_w =\frac{( M - N )}{M} $ where M is the total number of
configurations sampled in the production run and N is the lowest integer which
satisfies the condition

\begin{equation} 
  \frac{\sum_i^N w_i}{\sum_j^M w_j} \geqslant 0.5,
  \label{eq:iwcondition} 
\end{equation}

where the weights $w_i$ have been sorted in ascending order. In practice, the
$I_w$ index reflects how many configurations contribute after the reweighting in
\eq{eq:rew} to half of the total statistical weight. \\

  In the case of an extremely small overlap between configurational spaces
sampled by the production and target approximations, we would have $N\approx
M$ (since most weights $w_i$ would be close to 0, N has to approach M to
satisfy the condition in \eq{eq:iwcondition}) and $I_w \approx 0$. This
corresponds to a significant loss in the statistical significance since the
large number of configurations sampled in the production run is reduced to a few
individual structures. In the opposite case of excellent overlap we would have
$N\approx M/2$ and $I_w \approx 0.5$; if for example the production and target
approximations provides energies that differ only by a constant we would obtain
$I_w = 0.5$.\\

  The $I_w$ values for all systems investigated in this work are presented
in \tab{tab:Iw}; in order to simplify the analysis, we are also reporting
in parenthesis the deviations of the MLPT ensemble energies with respect to
the exact MD references (same as in \tab{tab:ergrew}). All the the $I_w$
values rounded to the value 0.00 actually correspond to values that are always
smaller than $0.003$. As expected the $I_w$ index is equal to 0.5 when the
production and target methods are the same. The values of $I_w$ allow for an
interpretation of the deviations previously reported in \tab{tab:ergrew}.
All the largest errors of MLPT (often significantly above chemical accuracy)
correspond to $I_w=0.00$, namely to a small correlation/overlap of production
and target methods. In few cases, small deviations are found for systems with
$I_w=0.00$ (for example the MLPT evaluation of the SCAN ensemble energy from
PBE+D2 for HCHAB) but these results must be considered as coincidental, whereby
the reweighting of a very small number of configurations led, by chance, to a
satisfactory reconstruction of the target statistical distribution.\\

  As discussed above, the MLPT estimates become reliable for the production and target
method combinations from within the two groups of functionals PBE/PBE+D2/vdW-DF2
and SCAN/SCAN+rVV10. As shown in \tab{tab:Iw} the $I_w$ values within this two
groups are all ``finite'' and take a value of at least 0.03. For the isolated
gas phase molecules \chfour and \cotwo we always find large values of $I_w$
($>$0.1) and consequently the errors on the energy predictions are small (within
0.18 kcal/mol). These remarks point to the fact that the $I_w$ can be used for a
qualitative \emph{a posteriori} estimate of the reliability of MLPT. In the case
of an $I_w$ close to 0, the MLPT estimates are unpredictable and often deviate
by several kcal/mol from the reference values. Non-vanishing values of $I_w$,
ideally as close as possible to 0.5, correspond to accurate MLPT predictions. In
the specific case of the systems considered here an $I_w$ of about 0.03 already
provides ensemble energies well within chemical accuracy.\\

\clearpage

\begin{table}[htpb!]
\footnotesize
\centering
\caption{\label{tab:Iw} Values of the $I_w$ index corresponding to
each individual MLPT estimate of the ensemble internal energy.
The values in parentheses represent the 
deviations in kcal/mol from the reference values obtained from straightforward MD.}
\begin{tabular}{ccccccc}
\hline
\hline
       &        & \multicolumn{5}{c}{Production - MD simulations} \\
\cline{3-7}
System & Target - MLPT & PBE & PBE+D2 & SCAN & vdW-DF2 & SCAN+rVV10 \\
\hline
\\
              &        PBE & \hres{ 0.50}{-0.01} & \hres{ 0.06}{ 0.05} & \hres{ 0.00}{-1.35} & \hres{ 0.03}{-0.24} & \hres{ 0.00}{-1.58}  \\
              &     PBE+D2 & \hres{ 0.03}{ 0.20} & \hres{ 0.50}{ 0.00} & \hres{ 0.00}{-0.55} & \hres{ 0.06}{ 0.41} & \hres{ 0.00}{-2.99}  \\
\chfour@HCHAB &       SCAN & \hres{ 0.00}{-4.38} & \hres{ 0.00}{-2.14} & \hres{ 0.50}{-0.03} & \hres{ 0.00}{ 1.50} & \hres{ 0.35}{ 0.20}  \\
              &    vdW-DF2 & \hres{ 0.03}{-0.38} & \hres{ 0.05}{ 0.21} & \hres{ 0.00}{ 0.28} & \hres{ 0.50}{ 0.06} & \hres{ 0.00}{-3.80}  \\
              & SCAN+rVV10 & \hres{ 0.00}{-5.03} & \hres{ 0.00}{-2.04} & \hres{ 0.34}{-0.22} & \hres{ 0.00}{ 1.01} & \hres{ 0.50}{-0.00}  \\
\\
              &        PBE & \hres{ 0.50}{ 0.01} & \hres{ 0.10}{-0.69} & \hres{ 0.00}{-2.95} & \hres{ 0.04}{-0.43} & \hres{ 0.00}{-3.69}  \\
              &     PBE+D2 & \hres{ 0.08}{ 0.24} & \hres{ 0.50}{ 0.04} & \hres{ 0.00}{-2.27} & \hres{ 0.12}{ 0.02} & \hres{ 0.00}{-1.67}  \\
\cotwo@SiCHAB &       SCAN & \hres{ 0.00}{ 2.25} & \hres{ 0.00}{ 1.24} & \hres{ 0.50}{-0.01} & \hres{ 0.00}{ 2.45} & \hres{ 0.35}{-0.59}  \\
              &    vdW-DF2 & \hres{ 0.03}{-0.05} & \hres{ 0.13}{-0.62} & \hres{ 0.00}{-4.14} & \hres{ 0.50}{-0.00} & \hres{ 0.00}{-2.12}  \\
              & SCAN+rVV10 & \hres{ 0.00}{ 3.39} & \hres{ 0.00}{ 2.29} & \hres{ 0.32}{ 0.90} & \hres{ 0.00}{ 3.25} & \hres{ 0.50}{ 0.02}  \\
\\
              &        PBE & \hres{ 0.50}{ 0.03} & \hres{ 0.31}{ 0.03} & \hres{ 0.00}{-3.71} & \hres{ 0.09}{-0.71} & \hres{ 0.00}{-2.95}  \\
              &     PBE+D2 & \hres{ 0.32}{ 0.05} & \hres{ 0.50}{ 0.00} & \hres{ 0.00}{-4.01} & \hres{ 0.11}{-0.63} & \hres{ 0.00}{-3.48}  \\
HCHAB         &       SCAN & \hres{ 0.00}{ 4.14} & \hres{ 0.00}{ 0.08} & \hres{ 0.50}{ 0.02} & \hres{ 0.00}{ 0.72} & \hres{ 0.39}{-0.03}  \\
              &    vdW-DF2 & \hres{ 0.08}{ 0.23} & \hres{ 0.11}{ 0.45} & \hres{ 0.00}{-4.41} & \hres{ 0.50}{-0.04} & \hres{ 0.00}{-3.68}  \\
              & SCAN+rVV10 & \hres{ 0.00}{ 3.83} & \hres{ 0.00}{-0.05} & \hres{ 0.39}{ 0.05} & \hres{ 0.00}{ 0.36} & \hres{ 0.50}{-0.03}  \\
\\
              &        PBE & \hres{ 0.50}{-0.01} & \hres{ 0.36}{-0.38} & \hres{ 0.00}{-2.63} & \hres{ 0.14}{-0.35} & \hres{ 0.00}{-0.23}  \\
              &     PBE+D2 & \hres{ 0.35}{ 0.36} & \hres{ 0.50}{ 0.00} & \hres{ 0.00}{-2.34} & \hres{ 0.14}{ 0.14} & \hres{ 0.00}{ 0.35}  \\
SiCHAB        &       SCAN & \hres{ 0.00}{ 0.66} & \hres{ 0.00}{ 1.05} & \hres{ 0.50}{ 0.00} & \hres{ 0.00}{ 3.81} & \hres{ 0.40}{ 0.14}  \\
              &    vdW-DF2 & \hres{ 0.13}{ 0.45} & \hres{ 0.15}{ 0.01} & \hres{ 0.00}{-2.97} & \hres{ 0.50}{ 0.00} & \hres{ 0.00}{-0.23}  \\
              & SCAN+rVV10 & \hres{ 0.00}{ 0.25} & \hres{ 0.00}{ 0.74} & \hres{ 0.40}{-0.16} & \hres{ 0.00}{ 3.36} & \hres{ 0.50}{ 0.00}  \\
\\
              &        PBE & \hres{ 0.50}{ 0.01} & \hres{ 0.46}{-0.05} & \hres{ 0.29}{-0.02} & \hres{ 0.33}{ 0.01} & \hres{ 0.28}{ 0.01}  \\
              &     PBE+D2 & \hres{ 0.46}{ 0.05} & \hres{ 0.50}{ 0.00} & \hres{ 0.26}{ 0.02} & \hres{ 0.29}{ 0.05} & \hres{ 0.25}{ 0.06}  \\
\chfour       &       SCAN & \hres{ 0.29}{ 0.01} & \hres{ 0.26}{-0.04} & \hres{ 0.50}{-0.00} & \hres{ 0.45}{ 0.03} & \hres{ 0.49}{ 0.01}  \\
              &    vdW-DF2 & \hres{ 0.33}{-0.01} & \hres{ 0.30}{-0.07} & \hres{ 0.45}{-0.03} & \hres{ 0.50}{-0.01} & \hres{ 0.44}{-0.01}  \\
              & SCAN+rVV10 & \hres{ 0.29}{ 0.00} & \hres{ 0.25}{-0.05} & \hres{ 0.49}{-0.01} & \hres{ 0.44}{ 0.01} & \hres{ 0.50}{-0.00}  \\
\\
              &        PBE & \hres{ 0.50}{ 0.00} & \hres{ 0.50}{-0.02} & \hres{ 0.12}{-0.03} & \hres{ 0.48}{-0.02} & \hres{ 0.14}{ 0.04}  \\
              &     PBE+D2 & \hres{ 0.50}{ 0.02} & \hres{ 0.50}{-0.00} & \hres{ 0.11}{-0.01} & \hres{ 0.48}{-0.01} & \hres{ 0.14}{ 0.06}  \\
\cotwo        &       SCAN & \hres{ 0.11}{ 0.05} & \hres{ 0.12}{-0.02} & \hres{ 0.50}{-0.01} & \hres{ 0.11}{-0.02} & \hres{ 0.45}{ 0.04}  \\
              &    vdW-DF2 & \hres{ 0.48}{ 0.03} & \hres{ 0.48}{ 0.01} & \hres{ 0.11}{ 0.00} & \hres{ 0.50}{-0.00} & \hres{ 0.13}{ 0.07}  \\
              & SCAN+rVV10 & \hres{ 0.14}{ 0.01} & \hres{ 0.15}{-0.05} & \hres{ 0.45}{-0.04} & \hres{ 0.14}{-0.05} & \hres{ 0.50}{ 0.00}  \\
\\
\hline
\hline
\end{tabular}
\end{table}

\clearpage

\subsection{Machine learning Monte Carlo resampling}

  Up to this point we have discussed the origin of MLPT inaccuracies occurring in
certain cases. Ideally, the production trajectory should be chosen to suit the
target approximations of interest. However, this is not always possible and it
is of interest to improve MLPT results also in the most pathological cases with
$I_w\approx 0$. Recently, Rizzi \etal~\cite{Rizzi2021} have discussed a ML-based
approach to overcome the limitations of TPT for free energy calculations in
cases where there is none or very limited overlap of the configurational
spaces of production and target methodologies. Following the previous approach
presented by Wirnsberger \etal~\cite{Wirnsberger2020}, 
Rizzi \etal~\cite{Rizzi2021} reported the
use of a configurational space transformation built on a neural network model.\\

  While this methodology is appealing, in this work we decided to pursue
an approach based on the Monte Carlo resampling of the trajectory. This
machine learning Monte Carlo (MLMC) method has the following advantages: The
implementation is straightforward and does not require the calculation of
forces, which often are not available in quantum chemical (i.e. RPA, MP2, or
CC) implementations available in solid state physics codes; the same ML model
of MLPT is reused and no additional calculations at the target level of theory
are required; the MC resampling can be used also to verify results when a good
or reasonable value of $I_w$ is found. Finally, let us remark that the kernel
methods used in our MLPT and MLMC methods typically require much smaller number
of training configurations than neural networks~\cite{Jablonka2020} used in the
method of Wirnsberger \etal~\cite{Wirnsberger2020}.\\

\begin{figure}[htpb!]
\includegraphics[width=16cm]{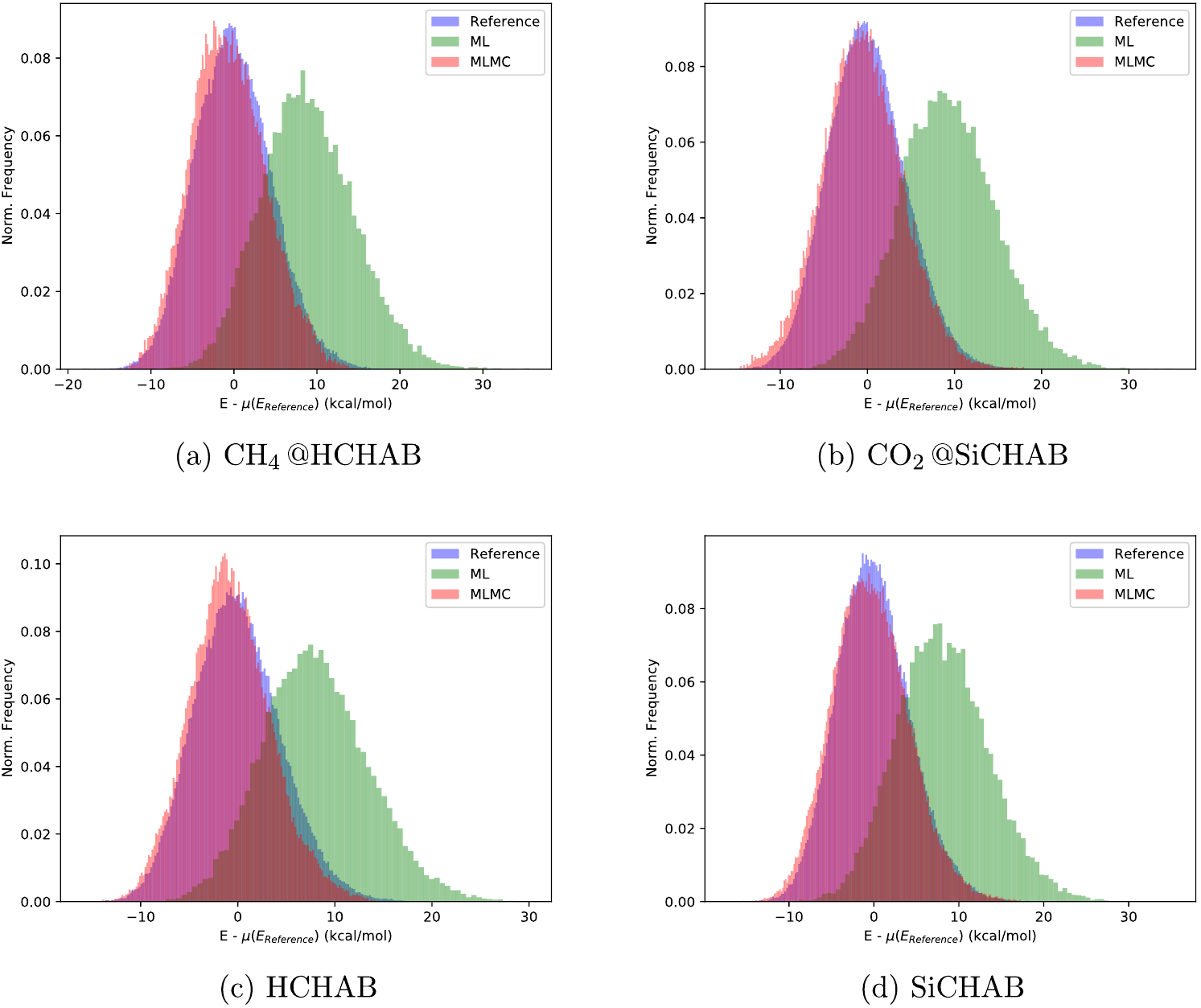}
\caption{ Distributions of deviations of SCAN energy from the average
value of the corresponding reference calculation ($\mu(E_{Reference})$) 
obtained in three different simulations: straightforward MD calculation 
with the SCAN functional (blue), ML (green) and MLMC (red) with PBE production
method.
}
\label{fig:hpsc}
\end{figure}

\begin{figure}[htpb!]
\includegraphics[width=16cm]{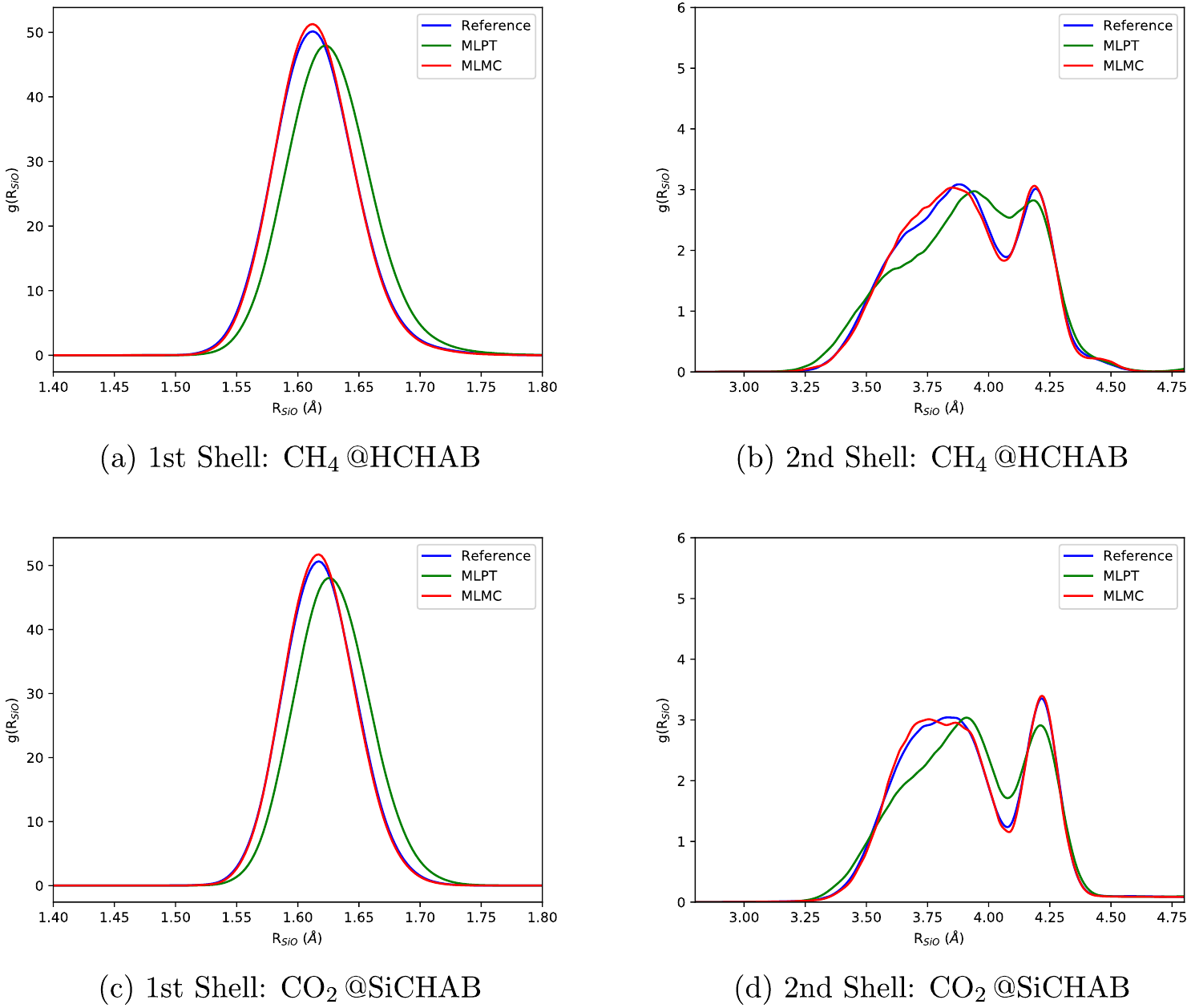}
\caption{Radial distribution functions calculated using the SCAN functional for 
the first and second coordination sphere of O atoms around Si. The calculations 
were performed using  a straightforward MD simulation performed at the SCAN level
(Reference), and via the MLPT and MLMC methods based on the PBE production method.}
\label{fig:rdfpsc}
\end{figure}

  To test the MLMC approach we consider the prediction of SCAN and SCAN+rVV10
enthalpies of adsorption from PBE production calculations which showed
the largest deviations between MLPT estimates and reference values. In
\fig{fig:hpsc} we show the following probability distributions of the SCAN
energy determined for each zeolite and adsorbed system: the exact reference from
the SCAN molecular dynamics (in blue labeled as reference); the distribution
of the ML predicted SCAN energies for the configurations sampled by the PBE
AIMD (denoted as ML in green); the distribution of SCAN energies from the MLMC
resampling (in red). The reference and ML distributions show an unsatisfactory
overlap, further confirming the unsatisfactory sampling of PBE to predict
SCAN properties. Since no reweighting of energies (such as in \eq{eq:rew})
has been performed to generate the histograms shown in \fig{fig:hpsc}, the ML
distributions are strongly biased, and therefore deviate significantly from
the reference straightforward MD results. The MLMC approach largely overcomes
this issue, as it can be noticed from the good overlap of the blue and red
curves. Computing the partial radial distribution function for the Si-O pairs,
\fig{fig:rdfpsc}, it is also evident that the MLMC generates geometries
which are in a better agreement with those obtained from straightforward MD
simulations with the target functional SCAN/SCAN+rVV10.\\

\begin{figure}[htpb!]
\centering
\includegraphics[width=16cm]{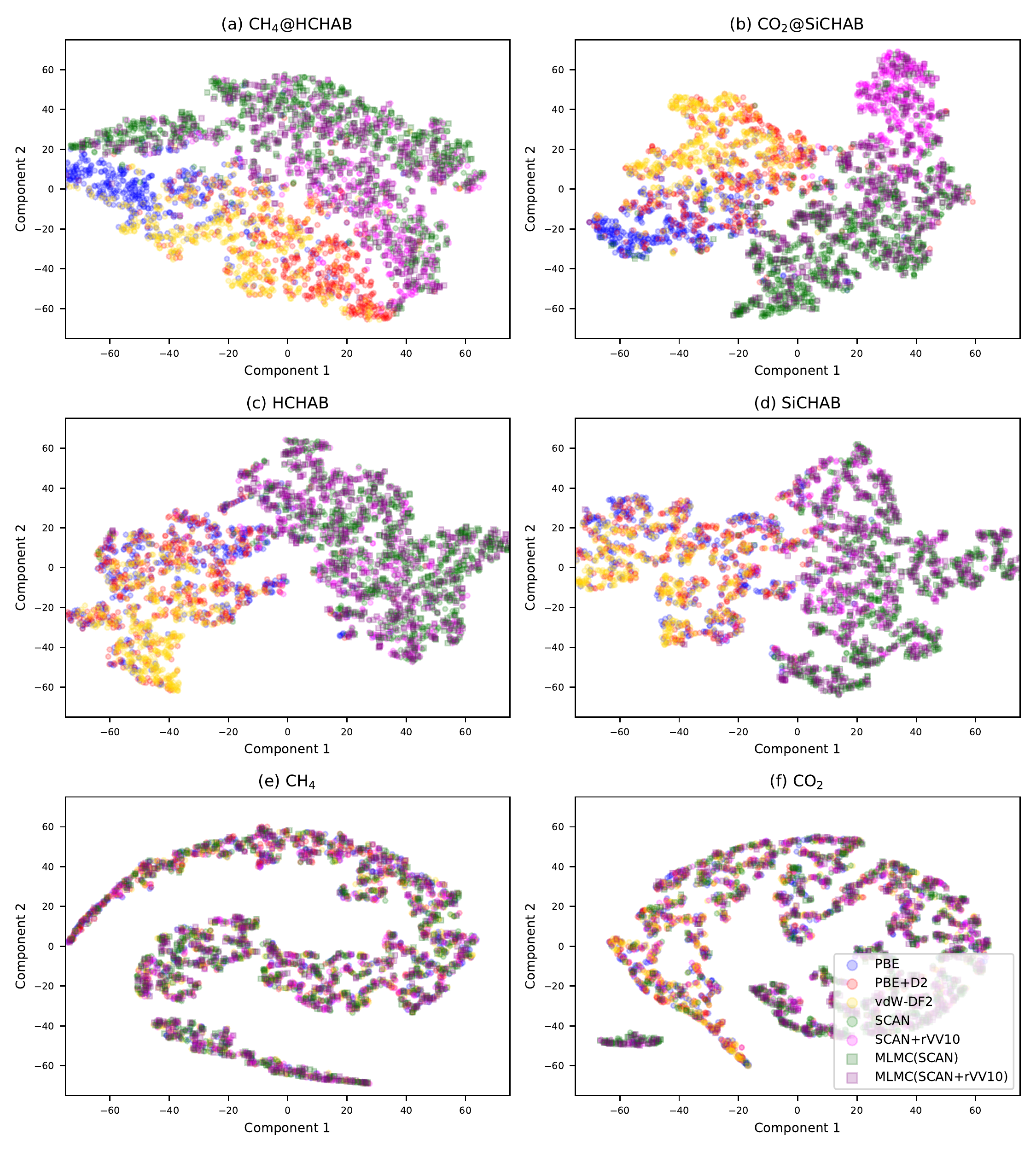}
\caption{t-SNE representation of the configurational spaces spanned
by the different functionals. This visualization is based on 500 selected 
configurations from each molecular dynamics trajectory and from
MLMC trajectories of the SCAN (MLMC(SCAN)) and SCAN+rVV10 (MLMC(SCAN+rVV10))
functionals using PBE as production method.}
\label{fig:mctsne}
\end{figure}

  The improvement in the geometric sampling can be also demonstrated by including
the MC data in the t-SNE analysis. \fig{fig:mctsne} shows indeed that
the clusters of the SCAN and SCAN+rVV10 configurations sampled by the Monte
Carlo have an excellent overlap with those obtained from the reference MD
simulation performed at the SCAN level. In this context it is important to
emphasize again that the MLMC results are obtained without any additional
explicit target level calculation.\\

\clearpage 

\clearpage 

\clearpage 

\clearpage

  We have shown insofar that the MLMC can significantly improve the sampling
of the geometries at the target level. We now show in \tab{tab:mcerr} the
improvement in the ensemble energies. The MLMC clearly reduces error in computed
internal energies of individual systems, as shown in \tab{tab:mcerr}. For
instance, the deviation in the internal energy of the \chfour@HCHAB at the SCAN
level was reduced from -4.38 kcal/mol from MLPT with PBE production method
to about -0.64 kcal/mol. Similarly, the deviations were also reduced for the
predicted SCAN and SCAN+rVV10 enthalpies of adsorptions, see \tab{tab:dHMC}.
For \chfour@HCHAB enthalpy, for instance, the MLPT presented a deviation of
-8.58 kcal/mol which reduced to only 0.12 kcal/mol when the MLMC resampling was
employed.\\

  Once again, the largest improvements have been observed for the adsorbate with
substrate systems, followed by the clean zeolites, while the effect of MLPT or
MLMC is negligible for the isolated molecular systems for which most of the
observed deviations are close to zero and probably are related to numerical
errors. In \mtab{tab:mcerr}{tab:dHMC}, all the predictions of SCAN/SCAN+rVV10
energies from PBE were significantly improved by the MLMC method. The general
improvement reduced the relative internal energy deviations below 1.0 kcal/mol.

\clearpage

\begin{table}[htpb!]
\footnotesize
\centering
\caption{\label{tab:mcerr} Deviations of the SCAN and SCAN+rVV10 internal
energies predicted using MLPT and MLMC
from their respective reference values obtained in straightforward MD 
simulations (\tab{tab:dHexp}).
All values are expressed in kcal/mol.}
\begin{tabular}{cccc}
\hline
\hline
\\
System & Prediction & MLPT &  MLMC \\
\hline
\\
\chfour@HCHAB &       SCAN & -4.380 & -0.639 \\
              & SCAN+rVV10 & -5.026 & -0.858 \\
\\
\cotwo@SiCHAB &       SCAN &  2.249 & -0.564 \\
              & SCAN+rVV10 &  3.395 &  0.961 \\
\\
HCHAB         &       SCAN &  4.142 & -0.749 \\
              & SCAN+rVV10 &  3.830 & -0.230 \\
\\
SiCHAB        &       SCAN &  0.659 & -0.345 \\
              & SCAN+rVV10 &  0.246 & -0.267 \\
\\
\chfour       &       SCAN &  0.013 & -0.011 \\
              & SCAN+rVV10 &  0.000 & -0.014 \\
\\
\cotwo        &       SCAN &  0.054 & -0.005 \\
              & SCAN+rVV10 &  0.013 & -0.058 \\
\\
\hline
\hline
\end{tabular}
\end{table}
\normalsize

\begin{table}[htpb!]
\footnotesize
\centering
\caption{\label{tab:dHMC} Experimental enthalpies of adsorption of
\chfour@HCHAB and \cotwo@SiCHAB and their calculated reference MLPT,
and MLMC values.
The deviations between the predictions and
the reference MD values are given in parentheses. All values are expressed in kcal/mol.}
\begin{tabular}{ccccc}
\hline
\hline
\\
System & Prediction & Ref. & MLPT  & MLMC \\
\hline
\\
\chfour@HCHAB                  &       SCAN & -2.60 & \hres{-11.13}{-8.54} & \hres{-2.48}{ 0.12} \\
Exp. -4.06~\cite{Piccini2015}   & SCAN+rVV10 & -4.77 & \hres{-13.63}{-8.86} & \hres{-5.38}{-0.61} \\
\\
\cotwo@SiCHAB                  &       SCAN & -2.95 & \hres{-1.42}{ 1.54} & \hres{-3.17}{-0.21} \\
Exp. -5.02~\cite{Maghsoudi2013} & SCAN+rVV10 & -7.11 & \hres{-3.97}{ 3.14} & \hres{-5.82}{ 1.29} \\
\\
\hline
\hline
\end{tabular}
\end{table}
\normalsize

\clearpage

\subsection{Assessing the accuracy of MLPT for the random phase approximation}

In a previous work (Ref.~\cite{Chehaibou2019}) some of us applied MLPT to compute
enthalpies of adsorption at the RPA level of theory. Beyond the two systems
studied in this work, also \chfour@CHAB and \cotwo@HCHAB were considered in
this previous paper. The RPA enthalpies of adsorption showed an excellent
agreement with experiment and for the molecules adsorbed in HCHAB zeolite a significant improvement
with respect to the production method (PBE+D2) was found (the results for
ensemble energies and enthalpies of adsorption will not be repeated here, where
we will rather discuss their reliability). Because of the high computational
cost involved in the RPA calculations in this case, it is not possible to produce
reference trajectories and use them to repeat some of the previous analysis
(indeed MLPT is meant to make possible finite-temperature calculations involving
computationally expensive approximations).\\

However, the $I_w$ index can be straightforwardly obtained and we present its
values in \tab{tab:Iwrpa}. This type of analysis, which we did not develop in
the original paper, Ref.~\cite{Chehaibou2019}, shows that a good performance of
MLPT should be expected, as the $I_w$ coefficient takes for all systems
values above 0.1. While not strictly necessary, the accuracy can be
further confirmed by applying MLMC, which requires several additional production
level calculations but avoids completely the highly expensive RPA calculations.
To this purpose we consider a subset of the systems in \tab{tab:Iwrpa} (the two adsorbed systems with the lowest $I_w$ and the two zeolites). With
respect to the MLPT values reported in Ref.~\cite{Chehaibou2019}, the MLMC
results present only small deviations: -0.21 kcal/mol for \cotwo@HCHAB, -0.48
kcal/mol for \chfour@HCHAB, -0.65 kcal/mol for HCHAB, and -0.05 kcal/mol for
SiCHAB. Keeping in mind that also statistical errors should be considered (the standard error
in the MC sampling is about 0.3 kcal/mol), we
conclude that the agreement between MLPT and MLMC results is very good from
which we deduce that PBE+D2 is a suitable production method for the RPA target
level calculations and that the Monte Carlo resampling of the MLPT results is
not strictly necessary.

\begin{table}[htpb!]
\footnotesize
\centering
\caption{\label{tab:Iwrpa} Values of the $I_w$ index corresponding to the
MLPT predictions of RPA energies from PBE+D2 trajectories for the different systems 
considered in Ref.~\cite{Chehaibou2019}.}
\begin{tabular}{cc}
\hline
\hline
   System       &  RPA  \\ 
\hline
\\
  \chfour@HCHAB &  0.13 \\ 
   \cotwo@HCHAB &  0.21 \\ 
 \chfour@SiCHAB &  0.26 \\ 
  \cotwo@SiCHAB &  0.27 \\ 
          HCHAB &  0.28 \\ 
         SiCHAB &  0.32 \\ 
\\
\hline
\hline
\end{tabular}
\end{table}

%
%

\clearpage

\section{Conclusions} \label{conclusions}

  We assessed the accuracy of the machine learning thermodynamic perturbation
theory approach by computing enthalpies of adsorption of molecules in zeolites.
By considering a set of five DFT functionals with different characteristics,
MLPT estimates are compared to reference results produced from full MD
simulations. The largest deviations are found when the configurational space
accessible to the target level has a small overlap with that sampled by the
production method. Even when an exact reference result is not available, the
most problematic cases can be detected by using a diagnostic test, the $I_w$
index. In these cases the results can be significantly improved by coupling the
Monte Carlo approach to the MLPT model to resample the configurational space
at the target level of theory which is, however, performed at the cost of the
production method.\\

  Finally, the analysis is extended to some recently published MLPT results that
used the RPA as target approximation~\cite{Chehaibou2019}; in this case the high
computational cost of the RPA makes the generation of exact reference ensemble
energies and enthalpies completely unpractical. The application of the $I_w$
diagnostic test and of the MC resampling highlights the full reliability of those
previous results, which were based on PBE+D2 production MD.\\

  The main advantage of the MLPT approach stands in the possibility of evaluating
finite temperature properties at a certain target level of theory by performing
only a minimal number of single-point calculations. Additionally, MLPT is also
suitable for approximations where the computation of gradients is not available
or adds a significant computational cost~\cite{ramberger17}. This approach opens
the possibility of systematically applying high-accuracy/high-cost \emph{ab
initio} methodologies to achieve a new level of predictive power in materials
simulations.

%
%

\clearpage

\begin{acknowledgement}
  This work was supported through the COMETE project (COnception in silico
de Mat\'eriaux pour l'EnvironnemenT et l'Energie) co-funded by the European 
Union under the program ``FEDER-FSE Lorraine et Massif des Vosges 2014-2020''.
This work was granted access to the HPC resources of TGCC under the 
allocations 2020- A0080810433 and 2021-A0100810433 by GENCI -EDARI project.
TB acknowledges support from the Slovak Research and Development Agency under 
the contracts  No.  VEGA-1/0777/19 and APVV-20-0127. 
\end{acknowledgement}

%
%

\clearpage 


\begin{mcitethebibliography}{72}
\providecommand*\natexlab[1]{#1}
\providecommand*\mciteSetBstSublistMode[1]{}
\providecommand*\mciteSetBstMaxWidthForm[2]{}
\providecommand*\mciteBstWouldAddEndPuncttrue
  {\def\EndOfBibitem{\unskip.}}
\providecommand*\mciteBstWouldAddEndPunctfalse
  {\let\EndOfBibitem\relax}
\providecommand*\mciteSetBstMidEndSepPunct[3]{}
\providecommand*\mciteSetBstSublistLabelBeginEnd[3]{}
\providecommand*\EndOfBibitem{}
\mciteSetBstSublistMode{f}
\mciteSetBstMaxWidthForm{subitem}{(\alph{mcitesubitemcount})}
\mciteSetBstSublistLabelBeginEnd
  {\mcitemaxwidthsubitemform\space}
  {\relax}
  {\relax}

\bibitem[M{\o}ller and Plesset(1934)M{\o}ller, and Plesset]{moller34}
M{\o}ller,~C.; Plesset,~M.~S. {Note on an Approximation Treatment for
  Many-Electron Systems}. \emph{Phys. Rev.} \textbf{1934}, \emph{46}, 618--622,
  DOI: \doi{10.1103/PhysRev.46.618}\relax
\mciteBstWouldAddEndPuncttrue
\mciteSetBstMidEndSepPunct{\mcitedefaultmidpunct}
{\mcitedefaultendpunct}{\mcitedefaultseppunct}\relax
\EndOfBibitem
\bibitem[Bartlett and Musia{\l}(2007)Bartlett, and Musia{\l}]{bartlett07}
Bartlett,~R.~J.; Musia{\l},~M. {Coupled-cluster theory in quantum chemistry}.
  \emph{Rev. Mod. Phys.} \textbf{2007}, \emph{79}, 291--352, DOI:
  \doi{10.1103/RevModPhys.79.291}\relax
\mciteBstWouldAddEndPuncttrue
\mciteSetBstMidEndSepPunct{\mcitedefaultmidpunct}
{\mcitedefaultendpunct}{\mcitedefaultseppunct}\relax
\EndOfBibitem
\bibitem[Pisani \latin{et~al.}(2008)Pisani, Maschio, Casassa, Halo,
  Sch{\"{u}}tz, and Usvyat]{pisani08}
Pisani,~C.; Maschio,~L.; Casassa,~S.; Halo,~M.; Sch{\"{u}}tz,~M.; Usvyat,~D.
  {Periodic local MP2 method for the study of electronic correlation in
  crystals: Theory and preliminary applications}. \emph{J. Comp. Chem.}
  \textbf{2008}, \emph{29}, 2113--2124, DOI: \doi{10.1002/jcc.20975}\relax
\mciteBstWouldAddEndPuncttrue
\mciteSetBstMidEndSepPunct{\mcitedefaultmidpunct}
{\mcitedefaultendpunct}{\mcitedefaultseppunct}\relax
\EndOfBibitem
\bibitem[Marsman \latin{et~al.}(2009)Marsman, Gr\"{u}neis, Paier, and
  Kresse]{marsman09}
Marsman,~M.; Gr\"{u}neis,~A.; Paier,~J.; Kresse,~G. {Second-order
  M{\o}ller–Plesset perturbation theory applied to extended systems. I.
  Within the projector-augmented-wave formalism using a plane wave basis set}.
  \emph{J. Chem. Phys.} \textbf{2009}, \emph{130}, 184103, DOI:
  \doi{10.1063/1.3126249}\relax
\mciteBstWouldAddEndPuncttrue
\mciteSetBstMidEndSepPunct{\mcitedefaultmidpunct}
{\mcitedefaultendpunct}{\mcitedefaultseppunct}\relax
\EndOfBibitem
\bibitem[{Del Ben} \latin{et~al.}(2012){Del Ben}, Hutter, and
  VandeVondele]{delben2012}
{Del Ben},~M.; Hutter,~J.; VandeVondele,~J. {Second-Order M{\o}ller–Plesset
  Perturbation Theory in the Condensed Phase: An Efficient and Massively
  Parallel Gaussian and Plane Waves Approach}. \emph{J. Chem. Theory Comput.}
  \textbf{2012}, \emph{8}, 4177--4188, DOI: \doi{10.1021/ct300531w}\relax
\mciteBstWouldAddEndPuncttrue
\mciteSetBstMidEndSepPunct{\mcitedefaultmidpunct}
{\mcitedefaultendpunct}{\mcitedefaultseppunct}\relax
\EndOfBibitem
\bibitem[Booth \latin{et~al.}(2013)Booth, Gr{\"{u}}neis, Kresse, and
  Alavi]{booth13}
Booth,~G.~H.; Gr{\"{u}}neis,~A.; Kresse,~G.; Alavi,~A. {Towards an exact
  description of electronic wavefunctions in real solids}. \emph{Nature}
  \textbf{2013}, \emph{493}, 365--370, DOI: \doi{10.1038/nature11770}\relax
\mciteBstWouldAddEndPuncttrue
\mciteSetBstMidEndSepPunct{\mcitedefaultmidpunct}
{\mcitedefaultendpunct}{\mcitedefaultseppunct}\relax
\EndOfBibitem
\bibitem[Dixit \latin{et~al.}(2017)Dixit, Claudot, Leb{\`{e}}gue, and
  Rocca]{dixit2017communication}
Dixit,~A.; Claudot,~J.; Leb{\`{e}}gue,~S.; Rocca,~D. {Communication: A novel
  implementation to compute MP2 correlation energies without basis set
  superposition errors and complete basis set extrapolation}. \emph{J. Chem.
  Phys.} \textbf{2017}, \emph{146}, 211102, DOI: \doi{10.1063/1.4985096}\relax
\mciteBstWouldAddEndPuncttrue
\mciteSetBstMidEndSepPunct{\mcitedefaultmidpunct}
{\mcitedefaultendpunct}{\mcitedefaultseppunct}\relax
\EndOfBibitem
\bibitem[Bohm and Pines(1953)Bohm, and Pines]{bohm53}
Bohm,~D.; Pines,~D. {A Collective Description of Electron Interactions: III.
  Coulomb Interactions in a Degenerate Electron Gas}. \emph{Phys. Rev.}
  \textbf{1953}, \emph{92}, 609--625, DOI: \doi{10.1103/PhysRev.92.609}\relax
\mciteBstWouldAddEndPuncttrue
\mciteSetBstMidEndSepPunct{\mcitedefaultmidpunct}
{\mcitedefaultendpunct}{\mcitedefaultseppunct}\relax
\EndOfBibitem
\bibitem[Gell-Mann and Brueckner(1957)Gell-Mann, and Brueckner]{gell57}
Gell-Mann,~M.; Brueckner,~K.~A. {Correlation Energy of an Electron Gas at High
  Density}. \emph{Phys. Rev.} \textbf{1957}, \emph{106}, 364--368, DOI:
  \doi{10.1103/PhysRev.106.364}\relax
\mciteBstWouldAddEndPuncttrue
\mciteSetBstMidEndSepPunct{\mcitedefaultmidpunct}
{\mcitedefaultendpunct}{\mcitedefaultseppunct}\relax
\EndOfBibitem
\bibitem[Langreth and Perdew(1975)Langreth, and Perdew]{langreth75}
Langreth,~D.; Perdew,~J. {The exchange-correlation energy of a metallic
  surface}. \emph{Solid State Commun.} \textbf{1975}, \emph{17}, 1425--1429,
  DOI: \doi{10.1016/0038-1098(75)90618-3}\relax
\mciteBstWouldAddEndPuncttrue
\mciteSetBstMidEndSepPunct{\mcitedefaultmidpunct}
{\mcitedefaultendpunct}{\mcitedefaultseppunct}\relax
\EndOfBibitem
\bibitem[Dobson and Wang(1999)Dobson, and Wang]{dobson99}
Dobson,~J.~F.; Wang,~J. {Successful Test of a Seamless van der Waals Density
  Functional}. \emph{Phys. Rev. Lett.} \textbf{1999}, \emph{82}, 2123--2126,
  DOI: \doi{10.1103/PhysRevLett.82.2123}\relax
\mciteBstWouldAddEndPuncttrue
\mciteSetBstMidEndSepPunct{\mcitedefaultmidpunct}
{\mcitedefaultendpunct}{\mcitedefaultseppunct}\relax
\EndOfBibitem
\bibitem[Furche(2001)]{furche01}
Furche,~F. {Molecular tests of the random phase approximation to the
  exchange-correlation energy functional}. \emph{Phys. Rev. B} \textbf{2001},
  \emph{64}, 195120, DOI: \doi{10.1103/PhysRevB.64.195120}\relax
\mciteBstWouldAddEndPuncttrue
\mciteSetBstMidEndSepPunct{\mcitedefaultmidpunct}
{\mcitedefaultendpunct}{\mcitedefaultseppunct}\relax
\EndOfBibitem
\bibitem[Harl and Kresse(2008)Harl, and Kresse]{harl08}
Harl,~J.; Kresse,~G. {Cohesive energy curves for noble gas solids calculated by
  adiabatic connection fluctuation-dissipation theory}. \emph{Phys. Rev. B}
  \textbf{2008}, \emph{77}, 045136, DOI: \doi{10.1103/PhysRevB.77.045136}\relax
\mciteBstWouldAddEndPuncttrue
\mciteSetBstMidEndSepPunct{\mcitedefaultmidpunct}
{\mcitedefaultendpunct}{\mcitedefaultseppunct}\relax
\EndOfBibitem
\bibitem[Lu \latin{et~al.}(2009)Lu, Li, Rocca, and Galli]{lu09}
Lu,~D.; Li,~Y.; Rocca,~D.; Galli,~G. {Ab initio Calculation of van der Waals
  Bonded Molecular Crystals}. \emph{Phys. Rev. Lett.} \textbf{2009},
  \emph{102}, 206411, DOI: \doi{10.1103/PhysRevLett.102.206411}\relax
\mciteBstWouldAddEndPuncttrue
\mciteSetBstMidEndSepPunct{\mcitedefaultmidpunct}
{\mcitedefaultendpunct}{\mcitedefaultseppunct}\relax
\EndOfBibitem
\bibitem[Gr{\"{u}}neis \latin{et~al.}(2009)Gr{\"{u}}neis, Marsman, Harl,
  Schimka, and Kresse]{gruneis09}
Gr{\"{u}}neis,~A.; Marsman,~M.; Harl,~J.; Schimka,~L.; Kresse,~G. {Making the
  random phase approximation to electronic correlation accurate}. \emph{J.
  Chem. Phys.} \textbf{2009}, \emph{131}, 154115, DOI:
  \doi{10.1063/1.3250347}\relax
\mciteBstWouldAddEndPuncttrue
\mciteSetBstMidEndSepPunct{\mcitedefaultmidpunct}
{\mcitedefaultendpunct}{\mcitedefaultseppunct}\relax
\EndOfBibitem
\bibitem[Olsen and Thygesen(2012)Olsen, and Thygesen]{olsen12}
Olsen,~T.; Thygesen,~K.~S. {Extending the random-phase approximation for
  electronic correlation energies: The renormalized adiabatic local density
  approximation}. \emph{Phys. Rev. B} \textbf{2012}, \emph{86}, 081103, DOI:
  \doi{10.1103/PhysRevB.86.081103}\relax
\mciteBstWouldAddEndPuncttrue
\mciteSetBstMidEndSepPunct{\mcitedefaultmidpunct}
{\mcitedefaultendpunct}{\mcitedefaultseppunct}\relax
\EndOfBibitem
\bibitem[Bates and Furche(2013)Bates, and Furche]{bates13}
Bates,~J.~E.; Furche,~F. {Communication: Random phase approximation
  renormalized many-body perturbation theory}. \emph{J. Chem. Phys.}
  \textbf{2013}, \emph{139}, 171103, DOI: \doi{10.1063/1.4827254}\relax
\mciteBstWouldAddEndPuncttrue
\mciteSetBstMidEndSepPunct{\mcitedefaultmidpunct}
{\mcitedefaultendpunct}{\mcitedefaultseppunct}\relax
\EndOfBibitem
\bibitem[Colonna \latin{et~al.}(2014)Colonna, Hellgren, and
  de~Gironcoli]{colonna14}
Colonna,~N.; Hellgren,~M.; de~Gironcoli,~S. {Correlation energy within
  exact-exchange adiabatic connection fluctuation-dissipation theory:
  Systematic development and simple approximations}. \emph{Phys. Rev. B}
  \textbf{2014}, \emph{90}, 125150, DOI: \doi{10.1103/PhysRevB.90.125150}\relax
\mciteBstWouldAddEndPuncttrue
\mciteSetBstMidEndSepPunct{\mcitedefaultmidpunct}
{\mcitedefaultendpunct}{\mcitedefaultseppunct}\relax
\EndOfBibitem
\bibitem[Dixit \latin{et~al.}(2016)Dixit, {\'{A}}ngy{\'{a}}n, and
  Rocca]{dixit16}
Dixit,~A.; {\'{A}}ngy{\'{a}}n,~J.~G.; Rocca,~D. {Improving the accuracy of
  ground-state correlation energies within a plane-wave basis set: The
  electron-hole exchange kernel}. \emph{J. Chem. Phys.} \textbf{2016},
  \emph{145}, 104105, DOI: \doi{10.1063/1.4962352}\relax
\mciteBstWouldAddEndPuncttrue
\mciteSetBstMidEndSepPunct{\mcitedefaultmidpunct}
{\mcitedefaultendpunct}{\mcitedefaultseppunct}\relax
\EndOfBibitem
\bibitem[Dixit \latin{et~al.}(2017)Dixit, Claudot, Leb{\`{e}}gue, and
  Rocca]{dixit17}
Dixit,~A.; Claudot,~J.; Leb{\`{e}}gue,~S.; Rocca,~D. {Improving the Efficiency
  of Beyond-RPA Methods within the Dielectric Matrix Formulation: Algorithms
  and Applications to the A24 and S22 Test Sets}. \emph{J. Chem. Theory
  Comput.} \textbf{2017}, \emph{13}, 5432--5442, DOI:
  \doi{10.1021/acs.jctc.7b00837}\relax
\mciteBstWouldAddEndPuncttrue
\mciteSetBstMidEndSepPunct{\mcitedefaultmidpunct}
{\mcitedefaultendpunct}{\mcitedefaultseppunct}\relax
\EndOfBibitem
\bibitem[Hellgren \latin{et~al.}(2018)Hellgren, Colonna, and
  de~Gironcoli]{hellgren18}
Hellgren,~M.; Colonna,~N.; de~Gironcoli,~S. {Beyond the random phase
  approximation with a local exchange vertex}. \emph{Phys. Rev. B}
  \textbf{2018}, \emph{98}, 045117, DOI: \doi{10.1103/PhysRevB.98.045117}\relax
\mciteBstWouldAddEndPuncttrue
\mciteSetBstMidEndSepPunct{\mcitedefaultmidpunct}
{\mcitedefaultendpunct}{\mcitedefaultseppunct}\relax
\EndOfBibitem
\bibitem[Behler and Parrinello(2007)Behler, and Parrinello]{behler07}
Behler,~J.; Parrinello,~M. {Generalized Neural-Network Representation of
  High-Dimensional Potential-Energy Surfaces}. \emph{Phys. Rev. Lett.}
  \textbf{2007}, \emph{98}, 146401, DOI:
  \doi{10.1103/PhysRevLett.98.146401}\relax
\mciteBstWouldAddEndPuncttrue
\mciteSetBstMidEndSepPunct{\mcitedefaultmidpunct}
{\mcitedefaultendpunct}{\mcitedefaultseppunct}\relax
\EndOfBibitem
\bibitem[Chmiela \latin{et~al.}(2017)Chmiela, Tkatchenko, Sauceda, Poltavsky,
  Sch{\"{u}}tt, and M{\"{u}}ller]{chmiela17}
Chmiela,~S.; Tkatchenko,~A.; Sauceda,~H.~E.; Poltavsky,~I.;
  Sch{\"{u}}tt,~K.~T.; M{\"{u}}ller,~K.-R. {Machine learning of accurate
  energy-conserving molecular force fields}. \emph{Sci. Adv.} \textbf{2017},
  \emph{3}, DOI: \doi{10.1126/sciadv.1603015}\relax
\mciteBstWouldAddEndPuncttrue
\mciteSetBstMidEndSepPunct{\mcitedefaultmidpunct}
{\mcitedefaultendpunct}{\mcitedefaultseppunct}\relax
\EndOfBibitem
\bibitem[Chmiela \latin{et~al.}(2018)Chmiela, Sauceda, M{\"{u}}ller, and
  Tkatchenko]{chmiela18}
Chmiela,~S.; Sauceda,~H.~E.; M{\"{u}}ller,~K.-R.; Tkatchenko,~A. {Towards exact
  molecular dynamics simulations with machine-learned force fields}. \emph{Nat.
  Commun.} \textbf{2018}, \emph{9}, 3887, DOI:
  \doi{10.1038/s41467-018-06169-2}\relax
\mciteBstWouldAddEndPuncttrue
\mciteSetBstMidEndSepPunct{\mcitedefaultmidpunct}
{\mcitedefaultendpunct}{\mcitedefaultseppunct}\relax
\EndOfBibitem
\bibitem[Zhang \latin{et~al.}(2018)Zhang, Han, Wang, Car, and E]{zhang18}
Zhang,~L.; Han,~J.; Wang,~H.; Car,~R.; E,~W. {Deep Potential Molecular
  Dynamics: A Scalable Model with the Accuracy of Quantum Mechanics}.
  \emph{Phys. Rev. Lett.} \textbf{2018}, \emph{120}, 143001, DOI:
  \doi{10.1103/PhysRevLett.120.143001}\relax
\mciteBstWouldAddEndPuncttrue
\mciteSetBstMidEndSepPunct{\mcitedefaultmidpunct}
{\mcitedefaultendpunct}{\mcitedefaultseppunct}\relax
\EndOfBibitem
\bibitem[Behler(2016)]{behler16}
Behler,~J. {Perspective: Machine learning potentials for atomistic
  simulations}. \emph{J. Chem. Phys.} \textbf{2016}, \emph{145}, 170901, DOI:
  \doi{10.1063/1.4966192}\relax
\mciteBstWouldAddEndPuncttrue
\mciteSetBstMidEndSepPunct{\mcitedefaultmidpunct}
{\mcitedefaultendpunct}{\mcitedefaultseppunct}\relax
\EndOfBibitem
\bibitem[Chehaibou \latin{et~al.}(2019)Chehaibou, Badawi, Bu{\v{c}}ko,
  Bazhirov, and Rocca]{Chehaibou2019}
Chehaibou,~B.; Badawi,~M.; Bu{\v{c}}ko,~T.; Bazhirov,~T.; Rocca,~D. {Computing
  RPA Adsorption Enthalpies by Machine Learning Thermodynamic Perturbation
  Theory}. \emph{J. Chem. Theory Comput.} \textbf{2019}, \emph{15}, 6333--6342,
  DOI: \doi{10.1021/acs.jctc.9b00782}\relax
\mciteBstWouldAddEndPuncttrue
\mciteSetBstMidEndSepPunct{\mcitedefaultmidpunct}
{\mcitedefaultendpunct}{\mcitedefaultseppunct}\relax
\EndOfBibitem
\bibitem[Bu{\v{c}}ko \latin{et~al.}(2020)Bu{\v{c}}ko,
  Ge{\v{s}}vandtnerov{\'{a}}, and Rocca]{Bucko2020}
Bu{\v{c}}ko,~T.; Ge{\v{s}}vandtnerov{\'{a}},~M.; Rocca,~D. {Ab Initio
  Calculations of Free Energy of Activation at Multiple Electronic Structure
  Levels Made Affordable: An Effective Combination of Perturbation Theory and
  Machine Learning}. \emph{J. Chem. Theory Comput.} \textbf{2020}, \emph{16},
  6049--6060, DOI: \doi{10.1021/acs.jctc.0c00486}\relax
\mciteBstWouldAddEndPuncttrue
\mciteSetBstMidEndSepPunct{\mcitedefaultmidpunct}
{\mcitedefaultendpunct}{\mcitedefaultseppunct}\relax
\EndOfBibitem
\bibitem[Ge{\v{s}}vandtnerov{\'{a}}
  \latin{et~al.}(2021)Ge{\v{s}}vandtnerov{\'{a}}, Rocca, and
  Bu{\v{c}}ko]{Gesvandtnerova2021}
Ge{\v{s}}vandtnerov{\'{a}},~M.; Rocca,~D.; Bu{\v{c}}ko,~T. {Methanol
  carbonylation over acid mordenite: Insights from ab initio molecular dynamics
  and machine learning thermodynamic perturbation theory}. \emph{J. Catal.}
  \textbf{2021}, \emph{396}, 166--178, DOI:
  \doi{10.1016/j.jcat.2021.02.011}\relax
\mciteBstWouldAddEndPuncttrue
\mciteSetBstMidEndSepPunct{\mcitedefaultmidpunct}
{\mcitedefaultendpunct}{\mcitedefaultseppunct}\relax
\EndOfBibitem
\bibitem[Ramberger \latin{et~al.}(2017)Ramberger, Sch{\"a}fer, and
  Kresse]{ramberger17}
Ramberger,~B.; Sch{\"a}fer,~T.; Kresse,~G. Analytic interatomic forces in the
  random phase approximation. \emph{Phys. Rev. Let.} \textbf{2017}, \emph{118},
  106403, DOI: \doi{10.1103/PhysRevLett.118.106403}\relax
\mciteBstWouldAddEndPuncttrue
\mciteSetBstMidEndSepPunct{\mcitedefaultmidpunct}
{\mcitedefaultendpunct}{\mcitedefaultseppunct}\relax
\EndOfBibitem
\bibitem[Grimme(2006)]{Grimme2006}
Grimme,~S. {Semiempirical GGA-type density functional constructed with a
  long-range dispersion correction}. \emph{J. Comp. Chem.} \textbf{2006},
  \emph{27}, 1787--1799, DOI: \doi{10.1002/jcc.20495}\relax
\mciteBstWouldAddEndPuncttrue
\mciteSetBstMidEndSepPunct{\mcitedefaultmidpunct}
{\mcitedefaultendpunct}{\mcitedefaultseppunct}\relax
\EndOfBibitem
\bibitem[Bučko \latin{et~al.}(2010)Bučko, Hafner, Lebègue, and
  Angyan]{Bucko2010}
Bučko,~T.; Hafner,~J.; Lebègue,~S.; Angyan,~J.~G. {Improved Description of
  the Structure of Molecular and Layered Crystals: Ab Initio DFT Calculations
  with van der Waals Corrections}. \emph{J. Phys. Chem. A} \textbf{2010},
  \emph{114}, 11814--11824, DOI: \doi{10.1021/jp106469x}\relax
\mciteBstWouldAddEndPuncttrue
\mciteSetBstMidEndSepPunct{\mcitedefaultmidpunct}
{\mcitedefaultendpunct}{\mcitedefaultseppunct}\relax
\EndOfBibitem
\bibitem[Ramakrishnan \latin{et~al.}(2015)Ramakrishnan, Dral, Rupp, and von
  Lilienfeld]{Ramakrishnan2015}
Ramakrishnan,~R.; Dral,~P.~O.; Rupp,~M.; von Lilienfeld,~O.~A. {Big Data Meets
  Quantum Chemistry Approximations: The $\Delta$-Machine Learning Approach}.
  \emph{J. Chem. Theory Comput.} \textbf{2015}, \emph{11}, 2087--2096, DOI:
  \doi{10.1021/acs.jctc.5b00099}\relax
\mciteBstWouldAddEndPuncttrue
\mciteSetBstMidEndSepPunct{\mcitedefaultmidpunct}
{\mcitedefaultendpunct}{\mcitedefaultseppunct}\relax
\EndOfBibitem
\bibitem[Chipot and Pohorille(2016)Chipot, and Pohorille]{chipot07}
Chipot,~C.; Pohorille,~A. \emph{Free energy calculations: Theory and
  Applications in Chemistry and Biology}; Springer, 2016\relax
\mciteBstWouldAddEndPuncttrue
\mciteSetBstMidEndSepPunct{\mcitedefaultmidpunct}
{\mcitedefaultendpunct}{\mcitedefaultseppunct}\relax
\EndOfBibitem
\bibitem[Pohorille \latin{et~al.}(2010)Pohorille, Jarzynski, and
  Chipot]{Pohorille2010}
Pohorille,~A.; Jarzynski,~C.; Chipot,~C. {Good Practices in Free-Energy
  Calculations}. \emph{J. Phys. Chem. B} \textbf{2010}, \emph{114},
  10235--10253, DOI: \doi{10.1021/jp102971x}\relax
\mciteBstWouldAddEndPuncttrue
\mciteSetBstMidEndSepPunct{\mcitedefaultmidpunct}
{\mcitedefaultendpunct}{\mcitedefaultseppunct}\relax
\EndOfBibitem
\bibitem[Rocca \latin{et~al.}(2019)Rocca, Dixit, Badawi, Leb{\`{e}}gue, Gould,
  and Bu{\v{c}}ko]{Rocca2019}
Rocca,~D.; Dixit,~A.; Badawi,~M.; Leb{\`{e}}gue,~S.; Gould,~T.; Bu{\v{c}}ko,~T.
  {Bridging molecular dynamics and correlated wave-function methods for
  accurate finite-temperature properties}. \emph{Phys. Rev. Mater.}
  \textbf{2019}, \emph{3}, 040801, DOI:
  \doi{10.1103/PhysRevMaterials.3.040801}\relax
\mciteBstWouldAddEndPuncttrue
\mciteSetBstMidEndSepPunct{\mcitedefaultmidpunct}
{\mcitedefaultendpunct}{\mcitedefaultseppunct}\relax
\EndOfBibitem
\bibitem[Adil \latin{et~al.}(2017)Adil, Belmabkhout, Pillai, Cadiau, Bhatt,
  Assen, Maurin, and Eddaoudi]{Adil2017}
Adil,~K.; Belmabkhout,~Y.; Pillai,~R.~S.; Cadiau,~A.; Bhatt,~P.~M.;
  Assen,~A.~H.; Maurin,~G.; Eddaoudi,~M. {Gas/vapour separation using
  ultra-microporous metal–organic frameworks: insights into the
  structure/separation relationship}. \emph{Chem. Soc. Rev} \textbf{2017},
  \emph{46}, 3402--3430, DOI: \doi{10.1039/C7CS00153C}\relax
\mciteBstWouldAddEndPuncttrue
\mciteSetBstMidEndSepPunct{\mcitedefaultmidpunct}
{\mcitedefaultendpunct}{\mcitedefaultseppunct}\relax
\EndOfBibitem
\bibitem[Khalil \latin{et~al.}(2020)Khalil, Jabraoui, Leb{\`{e}}gue, Kim,
  Aguilera, Thomas, Maug{\'{e}}, and Badawi]{Khalil2020}
Khalil,~I.; Jabraoui,~H.; Leb{\`{e}}gue,~S.; Kim,~W.~J.; Aguilera,~L.-J.;
  Thomas,~K.; Maug{\'{e}},~F.; Badawi,~M. {Biofuel purification: Coupling
  experimental and theoretical investigations for efficient separation of
  phenol from aromatics by zeolites}. \emph{Chem. Eng. J.} \textbf{2020},
  \emph{402}, 126264, DOI: \doi{10.1016/j.cej.2020.126264}\relax
\mciteBstWouldAddEndPuncttrue
\mciteSetBstMidEndSepPunct{\mcitedefaultmidpunct}
{\mcitedefaultendpunct}{\mcitedefaultseppunct}\relax
\EndOfBibitem
\bibitem[Chibani \latin{et~al.}(2016)Chibani, Chebbi, Leb{\`{e}}gue,
  Bu{\v{c}}ko, and Badawi]{Chibani2016}
Chibani,~S.; Chebbi,~M.; Leb{\`{e}}gue,~S.; Bu{\v{c}}ko,~T.; Badawi,~M. {A DFT
  investigation of the adsorption of iodine compounds and water in H-, Na-,
  Ag-, and Cu- mordenite}. \emph{J. Chem. Phys.} \textbf{2016}, \emph{144},
  244705, DOI: \doi{10.1063/1.4954659}\relax
\mciteBstWouldAddEndPuncttrue
\mciteSetBstMidEndSepPunct{\mcitedefaultmidpunct}
{\mcitedefaultendpunct}{\mcitedefaultseppunct}\relax
\EndOfBibitem
\bibitem[Hessou \latin{et~al.}(2021)Hessou, Jabraoui, Khalil, Dziurla, and
  Badawi]{Hessou2021}
Hessou,~E.~P.; Jabraoui,~H.; Khalil,~I.; Dziurla,~M.-A.; Badawi,~M. {Ab initio
  screening of zeolite Y formulations for efficient adsorption of thiophene in
  presence of benzene}. \emph{Appl. Surf. Sci.} \textbf{2021}, \emph{541},
  148515, DOI: \doi{10.1016/j.apsusc.2020.148515}\relax
\mciteBstWouldAddEndPuncttrue
\mciteSetBstMidEndSepPunct{\mcitedefaultmidpunct}
{\mcitedefaultendpunct}{\mcitedefaultseppunct}\relax
\EndOfBibitem
\bibitem[Perdew \latin{et~al.}(1996)Perdew, Burke, and Ernzerhof]{Perdew1996}
Perdew,~J.~P.; Burke,~K.; Ernzerhof,~M. {Generalized Gradient Approximation
  Made Simple}. \emph{Phys. Rev. Lett.} \textbf{1996}, \emph{77}, 3865--3868,
  DOI: \doi{10.1103/PhysRevLett.77.3865}\relax
\mciteBstWouldAddEndPuncttrue
\mciteSetBstMidEndSepPunct{\mcitedefaultmidpunct}
{\mcitedefaultendpunct}{\mcitedefaultseppunct}\relax
\EndOfBibitem
\bibitem[Sun \latin{et~al.}(2015)Sun, Ruzsinszky, and Perdew]{Sun2015}
Sun,~J.; Ruzsinszky,~A.; Perdew,~J. {Strongly Constrained and Appropriately
  Normed Semilocal Density Functional}. \emph{Phys. Rev. Lett.} \textbf{2015},
  \emph{115}, 036402, DOI: \doi{10.1103/PhysRevLett.115.036402}\relax
\mciteBstWouldAddEndPuncttrue
\mciteSetBstMidEndSepPunct{\mcitedefaultmidpunct}
{\mcitedefaultendpunct}{\mcitedefaultseppunct}\relax
\EndOfBibitem
\bibitem[Sun \latin{et~al.}(2016)Sun, Remsing, Zhang, Sun, Ruzsinszky, Peng,
  Yang, Paul, Waghmare, Wu, Klein, and Perdew]{Sun2016}
Sun,~J.; Remsing,~R.~C.; Zhang,~Y.; Sun,~Z.; Ruzsinszky,~A.; Peng,~H.;
  Yang,~Z.; Paul,~A.; Waghmare,~U.; Wu,~X.; Klein,~M.~L.; Perdew,~J.~P.
  {Accurate first-principles structures and energies of diversely bonded
  systems from an efficient density functional}. \emph{Nat. Chem.}
  \textbf{2016}, \emph{8}, 831--836, DOI: \doi{10.1038/nchem.2535}\relax
\mciteBstWouldAddEndPuncttrue
\mciteSetBstMidEndSepPunct{\mcitedefaultmidpunct}
{\mcitedefaultendpunct}{\mcitedefaultseppunct}\relax
\EndOfBibitem
\bibitem[Rom{\'{a}}n-P{\'{e}}rez and Soler(2009)Rom{\'{a}}n-P{\'{e}}rez, and
  Soler]{Roman-Perez2009}
Rom{\'{a}}n-P{\'{e}}rez,~G.; Soler,~J.~M. {Efficient Implementation of a van
  der Waals Density Functional: Application to Double-Wall Carbon Nanotubes}.
  \emph{Phys. Rev. Lett.} \textbf{2009}, \emph{103}, 096102, DOI:
  \doi{10.1103/PhysRevLett.103.096102}\relax
\mciteBstWouldAddEndPuncttrue
\mciteSetBstMidEndSepPunct{\mcitedefaultmidpunct}
{\mcitedefaultendpunct}{\mcitedefaultseppunct}\relax
\EndOfBibitem
\bibitem[Dion \latin{et~al.}(2004)Dion, Rydberg, Schr{\"{o}}der, Langreth, and
  Lundqvist]{Dion2004}
Dion,~M.; Rydberg,~H.; Schr{\"{o}}der,~E.; Langreth,~D.~C.; Lundqvist,~B.~I.
  {Van der Waals Density Functional for General Geometries}. \emph{Phys. Rev.
  Lett.} \textbf{2004}, \emph{92}, 246401, DOI:
  \doi{10.1103/PhysRevLett.92.246401}\relax
\mciteBstWouldAddEndPuncttrue
\mciteSetBstMidEndSepPunct{\mcitedefaultmidpunct}
{\mcitedefaultendpunct}{\mcitedefaultseppunct}\relax
\EndOfBibitem
\bibitem[Peng \latin{et~al.}(2016)Peng, Yang, Perdew, and Sun]{Peng2016}
Peng,~H.; Yang,~Z.-H.; Perdew,~J.~P.; Sun,~J. {Versatile van der Waals Density
  Functional Based on a Meta-Generalized Gradient Approximation}. \emph{Phys.
  Rev. X} \textbf{2016}, \emph{6}, 041005, DOI:
  \doi{10.1103/PhysRevX.6.041005}\relax
\mciteBstWouldAddEndPuncttrue
\mciteSetBstMidEndSepPunct{\mcitedefaultmidpunct}
{\mcitedefaultendpunct}{\mcitedefaultseppunct}\relax
\EndOfBibitem
\bibitem[Kresse and Hafner(1993)Kresse, and Hafner]{Kresse1993}
Kresse,~G.; Hafner,~J. {Ab initio molecular dynamics for liquid metals}.
  \emph{Phys. Rev. B} \textbf{1993}, \emph{47}, 558--561, DOI:
  \doi{10.1103/PhysRevB.47.558}\relax
\mciteBstWouldAddEndPuncttrue
\mciteSetBstMidEndSepPunct{\mcitedefaultmidpunct}
{\mcitedefaultendpunct}{\mcitedefaultseppunct}\relax
\EndOfBibitem
\bibitem[Kresse and Furthm{\"{u}}ller(1996)Kresse, and
  Furthm{\"{u}}ller]{Kresse1996}
Kresse,~G.; Furthm{\"{u}}ller,~J. {Efficient iterative schemes for ab initio
  total-energy calculations using a plane-wave basis set}. \emph{Phys. Rev. B}
  \textbf{1996}, \emph{54}, 11169--11186, DOI:
  \doi{10.1103/PhysRevB.54.11169}\relax
\mciteBstWouldAddEndPuncttrue
\mciteSetBstMidEndSepPunct{\mcitedefaultmidpunct}
{\mcitedefaultendpunct}{\mcitedefaultseppunct}\relax
\EndOfBibitem
\bibitem[Kresse and Furthm{\"{u}}ller(1996)Kresse, and
  Furthm{\"{u}}ller]{Kresse1996a}
Kresse,~G.; Furthm{\"{u}}ller,~J. {Efficiency of ab-initio total energy
  calculations for metals and semiconductors using a plane-wave basis set}.
  \emph{Comp. Mater. Sci.} \textbf{1996}, \emph{6}, 15--50, DOI:
  \doi{10.1016/0927-0256(96)00008-0}\relax
\mciteBstWouldAddEndPuncttrue
\mciteSetBstMidEndSepPunct{\mcitedefaultmidpunct}
{\mcitedefaultendpunct}{\mcitedefaultseppunct}\relax
\EndOfBibitem
\bibitem[Andersen(1980)]{Andersen1980}
Andersen,~H.~C. {Molecular dynamics simulations at constant pressure and/or
  temperature}. \emph{J. Chem. Phys.} \textbf{1980}, \emph{72}, 2384--2393,
  DOI: \doi{10.1063/1.439486}\relax
\mciteBstWouldAddEndPuncttrue
\mciteSetBstMidEndSepPunct{\mcitedefaultmidpunct}
{\mcitedefaultendpunct}{\mcitedefaultseppunct}\relax
\EndOfBibitem
\bibitem[Schiferl and Wallace(1985)Schiferl, and Wallace]{Schiferl1985}
Schiferl,~S.~K.; Wallace,~D.~C. {Statistical errors in molecular dynamics
  averages}. \emph{J. Chem. Phys.} \textbf{1985}, \emph{83}, 5203--5209, DOI:
  \doi{10.1063/1.449733}\relax
\mciteBstWouldAddEndPuncttrue
\mciteSetBstMidEndSepPunct{\mcitedefaultmidpunct}
{\mcitedefaultendpunct}{\mcitedefaultseppunct}\relax
\EndOfBibitem
\bibitem[Kresse and Joubert(1999)Kresse, and Joubert]{Kresse1999}
Kresse,~G.; Joubert,~D. {From ultrasoft pseudopotentials to the projector
  augmented-wave method}. \emph{Phys. Rev. B} \textbf{1999}, \emph{59},
  1758--1775, DOI: \doi{10.1103/PhysRevB.59.1758}\relax
\mciteBstWouldAddEndPuncttrue
\mciteSetBstMidEndSepPunct{\mcitedefaultmidpunct}
{\mcitedefaultendpunct}{\mcitedefaultseppunct}\relax
\EndOfBibitem
\bibitem[Deringer \latin{et~al.}(2021)Deringer, Bernstein, Cs{\'{a}}nyi, {Ben
  Mahmoud}, Ceriotti, Wilson, Drabold, and Elliott]{Deringer2021}
Deringer,~V.~L.; Bernstein,~N.; Cs{\'{a}}nyi,~G.; {Ben Mahmoud},~C.;
  Ceriotti,~M.; Wilson,~M.; Drabold,~D.~A.; Elliott,~S.~R. {Origins of
  structural and electronic transitions in disordered silicon}. \emph{Nature}
  \textbf{2021}, \emph{589}, 59--64, DOI:
  \doi{10.1038/s41586-020-03072-z}\relax
\mciteBstWouldAddEndPuncttrue
\mciteSetBstMidEndSepPunct{\mcitedefaultmidpunct}
{\mcitedefaultendpunct}{\mcitedefaultseppunct}\relax
\EndOfBibitem
\bibitem[Bart{\'{o}}k \latin{et~al.}(2013)Bart{\'{o}}k, Kondor, and
  Cs{\'{a}}nyi]{Bartok2013}
Bart{\'{o}}k,~A.~P.; Kondor,~R.; Cs{\'{a}}nyi,~G. {On representing chemical
  environments}. \emph{Phys. Rev. B} \textbf{2013}, \emph{87}, 184115, DOI:
  \doi{10.1103/PhysRevB.87.184115}\relax
\mciteBstWouldAddEndPuncttrue
\mciteSetBstMidEndSepPunct{\mcitedefaultmidpunct}
{\mcitedefaultendpunct}{\mcitedefaultseppunct}\relax
\EndOfBibitem
\bibitem[Himanen \latin{et~al.}(2020)Himanen, J{\"{a}}ger, Morooka, {Federici
  Canova}, Ranawat, Gao, Rinke, and Foster]{Himanen2020}
Himanen,~L.; J{\"{a}}ger,~M.~O.; Morooka,~E.~V.; {Federici Canova},~F.;
  Ranawat,~Y.~S.; Gao,~D.~Z.; Rinke,~P.; Foster,~A.~S. {DScribe: Library of
  descriptors for machine learning in materials science}. \emph{Comput. Phys.
  Commun.} \textbf{2020}, \emph{247}, 106949, DOI:
  \doi{10.1016/j.cpc.2019.106949}\relax
\mciteBstWouldAddEndPuncttrue
\mciteSetBstMidEndSepPunct{\mcitedefaultmidpunct}
{\mcitedefaultendpunct}{\mcitedefaultseppunct}\relax
\EndOfBibitem
\bibitem[Behler(2011)]{behler11}
Behler,~J. {Atom-centered symmetry functions for constructing high-dimensional
  neural network potentials}. \emph{J. Chem. Phys.} \textbf{2011}, \emph{134},
  074106, DOI: \doi{10.1063/1.3553717}\relax
\mciteBstWouldAddEndPuncttrue
\mciteSetBstMidEndSepPunct{\mcitedefaultmidpunct}
{\mcitedefaultendpunct}{\mcitedefaultseppunct}\relax
\EndOfBibitem
\bibitem[Huo and Rupp(2018)Huo, and Rupp]{huo17}
Huo,~H.; Rupp,~M. Unified Representation of Molecules and Crystals for Machine
  Learning. \emph{\href{https://arxiv.org/abs/1704.06439}{arXiv:1704.06439
  [physics.chem-ph]}} \textbf{2018}, \relax
\mciteBstWouldAddEndPunctfalse
\mciteSetBstMidEndSepPunct{\mcitedefaultmidpunct}
{}{\mcitedefaultseppunct}\relax
\EndOfBibitem
\bibitem[Casier \latin{et~al.}(2021)Casier, {Chagas da Silva}, Badawi, Pascale,
  Bu{\v{c}}ko, Leb{\`{e}}gue, and Rocca]{Casier2021}
Casier,~B.; {Chagas da Silva},~M.; Badawi,~M.; Pascale,~F.; Bu{\v{c}}ko,~T.;
  Leb{\`{e}}gue,~S.; Rocca,~D. {Hybrid localized graph kernel for machine
  learning energy‐related properties of molecules and solids}. \emph{J. Comp.
  Chem.} \textbf{2021}, \emph{42}, 1390--1401, DOI:
  \doi{10.1002/jcc.26550}\relax
\mciteBstWouldAddEndPuncttrue
\mciteSetBstMidEndSepPunct{\mcitedefaultmidpunct}
{\mcitedefaultendpunct}{\mcitedefaultseppunct}\relax
\EndOfBibitem
\bibitem[Stuke \latin{et~al.}(2019)Stuke, Todorovi{\'{c}}, Rupp, Kunkel, Ghosh,
  Himanen, and Rinke]{Stuke2019}
Stuke,~A.; Todorovi{\'{c}},~M.; Rupp,~M.; Kunkel,~C.; Ghosh,~K.; Himanen,~L.;
  Rinke,~P. {Chemical diversity in molecular orbital energy predictions with
  kernel ridge regression}. \emph{J. Chem. Phys.} \textbf{2019}, \emph{150},
  204121, DOI: \doi{10.1063/1.5086105}\relax
\mciteBstWouldAddEndPuncttrue
\mciteSetBstMidEndSepPunct{\mcitedefaultmidpunct}
{\mcitedefaultendpunct}{\mcitedefaultseppunct}\relax
\EndOfBibitem
\bibitem[Pedregosa \latin{et~al.}(2011)Pedregosa, Varoquaux, Gramfort, Michel,
  Thirion, Grisel, Blondel, Prettenhofer, Weiss, Dubourg, Vanderplas, Passos,
  Cournapeau, Brucher, Perrot, and Duchesnay]{Pedregosa}
Pedregosa,~F. \latin{et~al.}  {Scikit-learn: Machine Learning in Python}.
  \emph{J. Mach. Learn. Res.} \textbf{2011}, \emph{12}, 2825--2830\relax
\mciteBstWouldAddEndPuncttrue
\mciteSetBstMidEndSepPunct{\mcitedefaultmidpunct}
{\mcitedefaultendpunct}{\mcitedefaultseppunct}\relax
\EndOfBibitem
\bibitem[Paquet and Viktor(2015)Paquet, and Viktor]{Paquet2015}
Paquet,~E.; Viktor,~H.~L. {Molecular Dynamics, Monte Carlo Simulations, and
  Langevin Dynamics: A Computational Review}. \emph{Biomed Res. Int.}
  \textbf{2015}, \emph{2015}, 1--18, DOI: \doi{10.1155/2015/183918}\relax
\mciteBstWouldAddEndPuncttrue
\mciteSetBstMidEndSepPunct{\mcitedefaultmidpunct}
{\mcitedefaultendpunct}{\mcitedefaultseppunct}\relax
\EndOfBibitem
\bibitem[Kremer and Grest(1991)Kremer, and Grest]{Kremer1991}
Kremer,~K.; Grest,~G.~S. {Monte Carlo and molecular dynamics simulations of
  polymers}. \emph{Phys. Scr.} \textbf{1991}, \emph{T35}, 61--65, DOI:
  \doi{10.1088/0031-8949/1991/T35/013}\relax
\mciteBstWouldAddEndPuncttrue
\mciteSetBstMidEndSepPunct{\mcitedefaultmidpunct}
{\mcitedefaultendpunct}{\mcitedefaultseppunct}\relax
\EndOfBibitem
\bibitem[Meimaroglou and Kiparissides(2014)Meimaroglou, and
  Kiparissides]{Meimaroglou2014}
Meimaroglou,~D.; Kiparissides,~C. {Review of Monte Carlo Methods for the
  Prediction of Distributed Molecular and Morphological Polymer Properties}.
  \emph{Ind. Eng. Chem. Res.} \textbf{2014}, \emph{53}, 8963--8979, DOI:
  \doi{10.1021/ie4033044}\relax
\mciteBstWouldAddEndPuncttrue
\mciteSetBstMidEndSepPunct{\mcitedefaultmidpunct}
{\mcitedefaultendpunct}{\mcitedefaultseppunct}\relax
\EndOfBibitem
\bibitem[Betancourt(2019)]{Betancourt2019}
Betancourt,~M. {The Convergence of Markov Chain Monte Carlo Methods: From the
  Metropolis Method to Hamiltonian Monte Carlo}. \emph{Ann. Phys.}
  \textbf{2019}, \emph{531}, 1700214, DOI: \doi{10.1002/andp.201700214}\relax
\mciteBstWouldAddEndPuncttrue
\mciteSetBstMidEndSepPunct{\mcitedefaultmidpunct}
{\mcitedefaultendpunct}{\mcitedefaultseppunct}\relax
\EndOfBibitem
\bibitem[Piccini \latin{et~al.}(2015)Piccini, Alessio, Sauer, Zhi, Liu,
  Kolvenbach, Jentys, and Lercher]{Piccini2015}
Piccini,~G.; Alessio,~M.; Sauer,~J.; Zhi,~Y.; Liu,~Y.; Kolvenbach,~R.;
  Jentys,~A.; Lercher,~J.~A. {Accurate Adsorption Thermodynamics of Small
  Alkanes in Zeolites. Ab initio Theory and Experiment for H-Chabazite}.
  \emph{J. Phys. Chem. C} \textbf{2015}, \emph{119}, 6128--6137, DOI:
  \doi{10.1021/acs.jpcc.5b01739}\relax
\mciteBstWouldAddEndPuncttrue
\mciteSetBstMidEndSepPunct{\mcitedefaultmidpunct}
{\mcitedefaultendpunct}{\mcitedefaultseppunct}\relax
\EndOfBibitem
\bibitem[Maghsoudi \latin{et~al.}(2013)Maghsoudi, Soltanieh, Bozorgzadeh, and
  Mohamadalizadeh]{Maghsoudi2013}
Maghsoudi,~H.; Soltanieh,~M.; Bozorgzadeh,~H.; Mohamadalizadeh,~A. {Adsorption
  isotherms and ideal selectivities of hydrogen sulfide and carbon dioxide over
  methane for the Si-CHA zeolite: comparison of carbon dioxide and methane
  adsorption with the all-silica DD3R zeolite}. \emph{Adsorption}
  \textbf{2013}, \emph{19}, 1045--1053, DOI:
  \doi{10.1007/s10450-013-9528-1}\relax
\mciteBstWouldAddEndPuncttrue
\mciteSetBstMidEndSepPunct{\mcitedefaultmidpunct}
{\mcitedefaultendpunct}{\mcitedefaultseppunct}\relax
\EndOfBibitem
\bibitem[{Van Der Maaten} and Hinton(2008){Van Der Maaten}, and
  Hinton]{VanDerMaaten2008}
{Van Der Maaten},~L.; Hinton,~G. \emph{{Visualizing Data using t-SNE}}; 2008;
  Vol.~9; pp 2579--2605\relax
\mciteBstWouldAddEndPuncttrue
\mciteSetBstMidEndSepPunct{\mcitedefaultmidpunct}
{\mcitedefaultendpunct}{\mcitedefaultseppunct}\relax
\EndOfBibitem
\bibitem[De \latin{et~al.}(2016)De, Bart{\'{o}}k, Cs{\'{a}}nyi, and
  Ceriotti]{De2016}
De,~S.; Bart{\'{o}}k,~A.~P.; Cs{\'{a}}nyi,~G.; Ceriotti,~M. {Comparing
  molecules and solids across structural and alchemical space}. \emph{Phys.
  Chem. Chem. Phys.} \textbf{2016}, \emph{18}, 13754--13769, DOI:
  \doi{10.1039/C6CP00415F}\relax
\mciteBstWouldAddEndPuncttrue
\mciteSetBstMidEndSepPunct{\mcitedefaultmidpunct}
{\mcitedefaultendpunct}{\mcitedefaultseppunct}\relax
\EndOfBibitem
\bibitem[Rizzi \latin{et~al.}(2021)Rizzi, Carloni, and Parrinello]{Rizzi2021}
Rizzi,~A.; Carloni,~P.; Parrinello,~M. {Targeted Free Energy Perturbation
  Revisited: Accurate Free Energies from Mapped Reference Potentials}. \emph{J.
  Phys. Chem. Lett.} \textbf{2021}, 9449--9454, DOI:
  \doi{10.1021/acs.jpclett.1c02135}\relax
\mciteBstWouldAddEndPuncttrue
\mciteSetBstMidEndSepPunct{\mcitedefaultmidpunct}
{\mcitedefaultendpunct}{\mcitedefaultseppunct}\relax
\EndOfBibitem
\bibitem[Wirnsberger \latin{et~al.}(2020)Wirnsberger, Ballard, Papamakarios,
  Abercrombie, Racani{\`{e}}re, Pritzel, {Jimenez Rezende}, and
  Blundell]{Wirnsberger2020}
Wirnsberger,~P.; Ballard,~A.~J.; Papamakarios,~G.; Abercrombie,~S.;
  Racani{\`{e}}re,~S.; Pritzel,~A.; {Jimenez Rezende},~D.; Blundell,~C.
  {Targeted free energy estimation via learned mappings}. \emph{J. Chem. Phys.}
  \textbf{2020}, \emph{153}, 144112, DOI: \doi{10.1063/5.0018903}\relax
\mciteBstWouldAddEndPuncttrue
\mciteSetBstMidEndSepPunct{\mcitedefaultmidpunct}
{\mcitedefaultendpunct}{\mcitedefaultseppunct}\relax
\EndOfBibitem
\bibitem[Jablonka \latin{et~al.}(2020)Jablonka, Ongari, Moosavi, and
  Smit]{Jablonka2020}
Jablonka,~K.~M.; Ongari,~D.; Moosavi,~S.~M.; Smit,~B. {Big-Data Science in
  Porous Materials: Materials Genomics and Machine Learning}. \emph{Chem. Rev.}
  \textbf{2020}, \emph{120}, 8066--8129, DOI:
  \doi{10.1021/acs.chemrev.0c00004}\relax
\mciteBstWouldAddEndPuncttrue
\mciteSetBstMidEndSepPunct{\mcitedefaultmidpunct}
{\mcitedefaultendpunct}{\mcitedefaultseppunct}\relax
\EndOfBibitem
\end{mcitethebibliography}

%
%

\providecommand{\latin}[1]{#1}
\makeatletter
\providecommand{\doi}
  {\begingroup\let\do\@makeother\dospecials
  \catcode`\{=1 \catcode`\}=2 \doi@aux}
\providecommand{\doi@aux}[1]{\endgroup\texttt{#1}}
\makeatother
\providecommand*\mcitethebibliography{\thebibliography}
\csname @ifundefined\endcsname{endmcitethebibliography}
  {\let\endmcitethebibliography\endthebibliography}{}

\end{document}